\documentclass[11pt,letterpaper]{article}
\usepackage[utf8]{inputenc}
\usepackage[T1]{fontenc}
\usepackage[margin=1in]{geometry}
\usepackage{amsmath,amssymb,amsthm,mathtools}
\usepackage{enumitem}
\usepackage{booktabs}
\usepackage[dvipsnames]{xcolor}
\usepackage{graphicx}
\usepackage{placeins}
\usepackage[linesnumbered,ruled,vlined]{algorithm2e}
\usepackage[square,comma]{natbib}
\let\citep\citeyearpar
\usepackage[colorlinks,linkcolor=blue!70!black,citecolor=green!50!black,urlcolor=blue!70!black]{hyperref}
\usepackage[capitalize,nameinlink]{cleveref}

\newtheorem{theorem}{Theorem}[section]
\newtheorem{lemma}[theorem]{Lemma}

\newtheorem{corollary}[theorem]{Corollary}

\theoremstyle{definition}

\newtheorem{remark}[theorem]{Remark}

\newcommand{\OPT}{\mathrm{OPT}}

\newcommand{\RR}{\mathrm{RR}}
\newcommand{\cons}{\mathrm{cons}}
\newcommand{\rob}{\mathrm{rob}}
\newcommand{\eps}{\varepsilon}
\newcommand{\one}{\mathbf{1}}
\crefname{algocf}{Algorithm}{Algorithms}
\newcommand{\MRC}{\textup{MRC}}
\newcommand{\RT}{\textup{RT}}
\newcommand{\E}{\mathbb{E}}
\newcommand{\Prob}{\mathbb{P}}

\title{Pareto-optimal Non-Clairvoyant Scheduling with Predictions}
\author{
Ziyad Benomar\thanks{This work was conducted outside of Amazon.}\\
Amazon \\
Luxembourg, Luxembourg
\and
Vianney Perchet\\
CREST, ENSAE \& Criteo AI Lab \\
Palaiseau, France
}
\date{}

\begin{document}
\maketitle

\begin{abstract}
We study preemptive single-machine non-clairvoyant scheduling with predictions, where a prediction of the order of the jobs (or of their sizes) is available, and the objective is to minimize the total completion time. An algorithm is $c$-consistent if it is $c$-competitive when the prediction is correct, and $R$-robust if it is $R$-competitive whatever the prediction. Since the introduction of this problem as one of the first examples of learning-augmented algorithms, the best possible trade-off between consistency and robustness has remained open. We determine it exactly for deterministic algorithms. For every $R>2$, we give an $R$-robust algorithm whose consistency is
\[
c(R)=R\Big(1-\Big(1-\frac{R+2}{R^2}\Big)^{\frac R{R+2}}\Big),
\]
and we prove that no deterministic $R$-robust algorithm, with either predicted order or predicted job sizes, has better consistency. In particular, consistency $1+\eps$ is achievable with robustness $\eps^{-1/2}+\frac16+O(\sqrt\eps)$, whereas all previous algorithms required robustness $\Theta(1/\eps)$. Our algorithm is greedy: it follows the predicted order as fast as possible, subject to the invariant that the cost paid so far is at most $R$ times the optimum of the instance in which every unfinished job would end immediately. The matching lower bound rests on the observation that every robust deterministic algorithm must satisfy the same invariant along its run on correct predictions. Furthermore, we give a simple randomized algorithm, using predicted sizes, with consistency $1+\eps$ and robustness $(3\eps)^{-1/2}+O(1)$, strictly below the deterministic frontier. The question of determining the Pareto-optimal frontier among randomized algorithms remains open.
\end{abstract}

\setcounter{tocdepth}{2}
\section{Introduction}\label{sec:intro}

Non-clairvoyant scheduling is a classical problem in online optimization and competitive analysis~\citep{BEY98}, with applications in operating systems and data centers. In the single-machine setting, $n$ jobs with unknown processing times $p_1, \ldots, p_n$ are given, and the goal is to schedule them preemptively to minimize the sum of their completion times $\sum_j C_j$. Jobs can be interrupted and resumed later, and the machine can be shared among several jobs. If the processing times are known, the optimal solution is to run the jobs in increasing order of their sizes. This rule is known as Shortest Processing Time first (SPT), or Smith's rule~\citep{Smith56}. In the non-clairvoyant setting, however, the scheduler only learns the processing time of a job when it completes. The objective is then to have an algorithm that minimizes the \emph{competitive ratio}~\citep{BEY98}, which is the worst-case ratio between the cost of the online algorithm and that of SPT. The optimal algorithm in this setting is Round-Robin (RR), which allocates the same amount of computing power to all unfinished jobs, and it is $2$-competitive~\citep{MPT94}.

The factor $2$ is a worst-case guarantee, and it is overly pessimistic in many practical scenarios where some information about the jobs is available. For example, job sizes can be estimated by machine learning models trained on past executions. This setting is captured by the framework of \emph{learning-augmented algorithms} (aka \emph{algorithms with predictions})~\citep{LV21,PSK18,MV22}, where the algorithm receives a prediction without any guarantee on its accuracy. An algorithm is \emph{$c$-consistent} if it is $c$-competitive when the prediction is perfect, and \emph{$R$-robust} if it is $R$-competitive for arbitrarily bad predictions. Blindly following the prediction yields a consistency of $1$ but an arbitrarily bad robustness, whereas ignoring it yields a robustness of $2$ without any improvement when the prediction is accurate. The goal is therefore to understand the trade-off between these two criteria. More precisely, for any consistency $c$, what is the best robustness $R^*(c)$ achievable by a non-clairvoyant algorithm? This defines the Pareto frontier of the problem.

Non-clairvoyant scheduling was studied in the learning-augmented framework from its start. \citet{PSK18} considered predictions of the job sizes, and showed that running Round-Robin and the schedule suggested by the predictions in parallel, at suitable rates, yields a $c$-consistent and $\frac{2c-1}{c-1}$-robust algorithm~\citep[Theorem~3.4]{PSK18}. \citet{LM22} later introduced a weaker prediction model, where the algorithm only receives a predicted order of the jobs. \citet{WZ20} established a lower bound on the trade-off, showing that any $c$-consistent algorithm has robustness at least $1/\sqrt{2(c-1)}$. They also characterized the exact trade-off for two jobs. The best known algorithm~\citep{BCLS25} achieves a robustness of $\frac{c}{c-1}$. Therefore, for a consistency of $c=1+\eps$, the known lower bound on robustness is $\Theta(\eps^{-1/2})$, and the best known algorithm is $\Theta(1/\eps)$-robust. The exact Pareto frontier for more than two jobs remained open.

\subsection{Contributions}\label{sec:contrib}

We determine in this paper the exact consistency--robustness frontier of deterministic non-clairvoyant scheduling with predictions, for both predicted orders and predicted job sizes. Moreover, we give a simple randomized algorithm with size predictions that beats the deterministic Pareto-optimal frontier.

We introduce a new algorithm \emph{Maximal Robust Commitment} ($\MRC_R$, \cref{alg:ft}), which only uses the predicted order. At any time $t$, it greedily keeps the cost paid so far within a factor $R$ of what the optimum would be if every unfinished job ended exactly at time $t$, which immediately gives robustness $R$. Subject to this invariant, it \emph{commits} as much as possible to the job that the prediction says to run next: it runs this job alone whenever this preserves the invariant, and otherwise runs it together with Round-Robin on the other jobs, with processing rates that keep the invariant tight. 
Our main result is a characterization of the consistency and robustness of $\MRC_R$, followed by a matching lower bound for any deterministic algorithm using either predicted order or sizes.

\begin{theorem}[Upper bound]\label{thm:ub}
For every $R>2$, $\MRC_R$ is $R$-robust and $c(R)$-consistent, where
\begin{equation}\label{eq:cR}
c(R)=R\Big(1-\Big(1-\frac{R+2}{R^2}\Big)^{\frac{R}{R+2}}\Big).
\end{equation}
\end{theorem}

\begin{theorem}[Lower bound]\label{thm:lbm}
For every $R>2$, any deterministic $R$-robust algorithm has consistency at least $c(R)$, whether it receives predicted orders or predicted sizes.
\end{theorem}

\begin{figure}[t]
\centering
\includegraphics{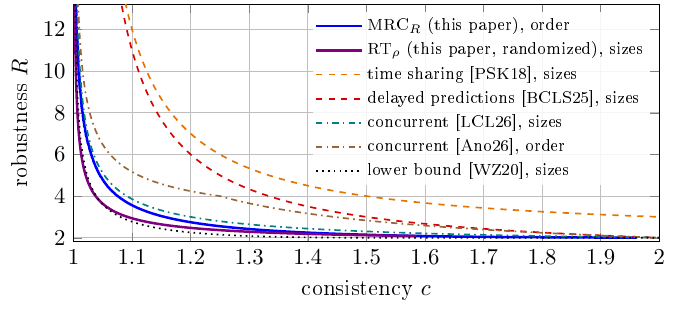}
\caption{Consistency--robustness trade-offs for non-clairvoyant scheduling. Upper bounds (deterministic algorithms): time sharing between the predicted schedule and Round-Robin~\citep{PSK18}, robustness $\frac{2c-1}{c-1}$; the delayed-predictions algorithm~\citep{BCLS25}, robustness $\frac c{c-1}$; the algorithms of the independent concurrent works~\citep{LCL26,Ano26} (dash-dotted lines); and $\MRC_R$, which attains the optimal deterministic trade-off $(c(R),R)$ (\cref{thm:ub,thm:lbm}). The dotted line is the lower bound of~\citep{WZ20}, maximized over the number of jobs. The randomized algorithm $\RT_\rho$ (\cref{thm:rand}) lies below the deterministic frontier. The legend states, for each curve, whether the algorithm uses predicted sizes or a predicted order.}
\label{fig:frontier}
\end{figure}

\Cref{thm:ub} is proved in \cref{app:run,app:ft,app:cons} (\cref{thm:ft-rob,thm:ft-cons}) and \cref{thm:lbm} in \cref{app:lbsec} (\cref{thm:lb}). Thus $c(R)$ is the best consistency of a deterministic $R$-robust algorithm, and it is attained by a greedy algorithm. Since the lower bound holds even with exact predicted sizes and the upper bound uses only their order, predicted sizes are worth exactly as much as a predicted order for this trade-off. The lower bound follows from the fact that every deterministic $R$-robust algorithm must satisfy the robustness invariant of $\MRC_R$ at every moment of its run on correct predictions (\cref{lem:freeze-lb}), and $\MRC_R$ is the algorithm that saturates this constraint. This extends the adversary argument of \citet{WZ20} from the first completion to every moment of the run. \Cref{fig:frontier} compares the obtained consistency--robustness trade-off with previous results. Writing $R^*(c_0)$ for the smallest robustness of a deterministic algorithm with consistency $c_0$, its behavior near consistency $1$ is (\cref{cor:frontier})
\[
R^*(1+\eps)=\eps^{-1/2}+\frac16+O(\sqrt\eps),
\]
a quadratic improvement over the robustness $\Theta(1/\eps)$ of previous algorithms. In simulations (\cref{sec:exp}), $\MRC_R$ is close to following the predictions when they are accurate, and is then far better than previous algorithms with the same robustness.

Without predictions, randomization does not help, since the optimal competitive ratio $2$ is attained by Round-Robin, which is deterministic. With predicted sizes, we prove that this is no longer the case, by giving a simple randomized algorithm whose trade-off is strictly below the deterministic frontier. For a randomized algorithm, both criteria are defined with the expected total completion time, against an oblivious adversary (\cref{sec:prelim}). Our \emph{randomized two-stage} algorithm $\RT_\rho$ (\cref{alg:rt}) runs Round-Robin until the common elapsed time of the jobs reaches a random level, then runs the jobs one at a time in the predicted order, and switches back to Round-Robin, among the jobs of smallest elapsed time, at the first job whose size contradicts its prediction.

\begin{theorem}[Randomization]\label{thm:rand}
For every $\rho\ge\rho_0=\frac{\sqrt{10}-3}2\approx0.081$, $\RT_\rho$ is $(1+\eps(\rho))$-consistent and $(2+\rho)$-robust, where $\eps(\rho)=\int_0^1\frac{v^2\,dv}{v^2+3\rho v+\rho^2}$, and $2+\rho<R^*(1+\eps(\rho))$. At consistency $1+\eps$, its robustness is $(3\eps)^{-1/2}+O(1)$ as $\eps\to0$.
\end{theorem}

For every consistency $c\in(1,1.585]$, $\RT_\rho$ with $\eps(\rho)=c-1$ has a robustness strictly below that of every deterministic $c$-consistent algorithm, and near consistency $1$ it improves the leading constant by a factor $\sqrt3$ (analysis in \cref{sec:rand-overview,app:rand}). Whether randomization helps with a predicted order only remains an open question.

\paragraph{Organization.} \Cref{sec:prelim} introduces the model and the two quantities that drive the analysis. \Cref{sec:algo} presents $\MRC_R$ and its robustness. \Cref{sec:overview} gives an overview of the proofs of consistency and of the lower bound, and of the analysis of $\RT_\rho$. \Cref{sec:discussion} discusses the frontier and open problems. All proofs are in the appendix, and \cref{sec:exp} reports experiments.

\subsection{Technical overview}\label{sec:challenges}

\paragraph{The robustness invariant.} A robust algorithm must \emph{hedge} against predictions that are wrong, that is, process jobs before their turn in the predicted order. How much hedging is necessary? Time sharing~\citep{PSK18} hedges at a fixed rate, independently of the state of the schedule, which is wasteful. Our starting point is the \emph{robustness invariant}: the cost paid so far stays within a factor $R$ of the optimum of the instance in which every unfinished job ends now. It is necessary for $R$-robustness, because at any time $t$ an adversary can declare that every unfinished job ends right after $t$, and a deterministic algorithm cannot distinguish this instance from the current one before time $t$. This adversary is the one used by \citet{WZ20} at the first completion. The invariant is also sufficient, and it quantifies how much hedging is needed at every moment.

\paragraph{Algorithm analysis.} The robustness of $\MRC_R$ follows directly from its definition. Consistency is harder to analyze. The run of $\MRC_R$ is explicit, but on a general instance whether it commits or hedges at a given moment depends on the sizes of all completed jobs. We decompose the cost into contributions of \emph{layers} of elapsed time. On a correct order, a layer costs more when more jobs are still unfinished as the schedule passes it. We compare the run with a \emph{target}, which prescribes, for every layer, the fraction of jobs that may still be unfinished at that point, and which is \emph{admissible} if a schedule following it would satisfy the invariant. A first-violation argument shows that $\MRC_R$ is never behind any admissible target. The consistency analysis then reduces to finding the best admissible target, a one-dimensional problem that is the same for all instances, and whose solution yields the consistency bound $c(R)$.

\paragraph{A matching lower bound.} Our lower bound extends the argument of \citet{WZ20}. On $n$ equal jobs, the robustness invariant yields one constraint at each of the $n$ completion times. Wei and Zhang use the first of them, which gives the right order $\eps^{-1/2}$ but not the exact frontier, since an algorithm can spread its hedging between completions. We use all $n$ constraints. Once the concave optimum $\Phi$ of the instance in which all unfinished jobs end immediately is bounded by a linear function, each constraint says that the hedging forced at a completion grows with the hedging already done. This gives a simple recursion, which is a discrete version of the differential equation that determines the optimal target. Solving it gives a bound on the excess cost which, divided by $\OPT$, converges to $c(R)-1$.

\subsection{Related work}\label{sec:related}

\paragraph{Learning-augmented algorithms.} Since the work of \citet{LV21} on caching and of \citet{PSK18} on ski rental and scheduling, the framework of algorithms with predictions has been applied to many online problems, including caching~\citep{LV21,Wei20}, metrical task systems~\citep{ACEPS23}, online primal-dual algorithms~\citep{BMS20}, speed scaling~\citep{BMRS20,AJS22}, online search~\citep{Ang23, BP25}, knapsack~\citep{IKQP21k}, online allocation problems~\citep{MSVV07,BJN07}, and matching~\citep{JM22}; see~\citep{MV22} for a survey. Other lines of work within this literature study multiple predictors~\citep{DILMV22} and learnable predictors~\citep{EKMM24}.

\paragraph{Scheduling with predictions.} After~\citep{PSK18}, non-clairvoyant scheduling with predictions has been studied with predicted sizes and refined error measures~\citep{IKQP21}, with predicted orders~\citep{LM22}, with untrusted predictions in further scheduling models~\citep{BDKLP22}, with predictions available for only some of the jobs~\citep{BP24}, with progress bars that reveal partial information during processing~\citep{BCLS25}, and with stochastic signals~\citep{CLLS26}. Related models include flow time with uncertain processing times~\citep{ALT21,ALT22}, jobs with online precedence constraints~\citep{LLMS23}, queueing with predicted service times~\citep{Mit20}, and online makespan scheduling with learned weights~\citep{LLMV20}. All previously known algorithms for our problem achieve robustness $\Theta(1/\eps)$ for consistency $1+\eps$, and the algorithm of~\citep{BCLS25} is the best known. The only exceptions are the algorithms of the concurrent works discussed below.

\paragraph{Optimal consistency--robustness trade-offs.} Pareto frontiers are known for only a few problems, among them ski rental~\citep{WZ20,ADJKR20}, online conversion and one-way trading~\citep{SLHWT21}, online search~\citep{LSHL24}, two-stage matching~\citep{JM22} and metrical task systems~\citep{CSW23}; near-optimal trade-offs are known for online knapsack~\citep{DKLSMH25}. \citet{WZ20} studied non-clairvoyant scheduling and solved the case of two jobs. We solve the general case of any number of jobs.

\paragraph{Independent concurrent work.} This paper was written independently of, and without knowledge of, two concurrent works, which appeared very shortly before it and which we became aware of after its completion. \citet{LCL26} use predicted sizes. For every number of jobs $n$, they prove a lower bound whose limit as $n\to\infty$ coincides with \cref{thm:lbm}, and give an algorithm whose excess consistency is within a factor $\frac32$ of this bound, and which, at consistency $1+\eps$, has robustness at most $\frac12+\sqrt{\frac54+\frac1\eps}$ for every $n$. This algorithm returns to Round-Robin at the first prediction error, so it is \emph{brittle} in the sense of~\citep{EADL24,BP25}: an arbitrarily small error on one predicted size raises its cost from the consistency bound to that of Round-Robin. In practice, predicted sizes can never be perfectly accurate, so the algorithm would behave as Round-Robin in any realistic scenario. \citet{Ano26} use a predicted order, and give a $(1+\eps)$-consistent algorithm that is $\min\{2\eps^{-1/2},2+\eps^{-1/2}\}$-robust. For instance, at robustness $3$ its consistency guarantee is $1+\frac49\approx1.44$, whereas $\MRC_3$ is $c(3)\approx1.16$-consistent. Both works close the quadratic gap, but neither algorithm is Pareto-optimal (\cref{fig:frontier,app:exp}). We determine the Pareto-optimal frontier exactly, attain it with a predicted order alone, and show that randomization beats it. Moreover, experiments show that our algorithm on natural instances outperforms the concurrent algorithms. On the other hand, \citet{LCL26} also give bounds for each fixed $n$, whereas ours are worst cases over all $n$, and for a fixed $n$ our proof only gives a lower bound (\cref{lem:lb-rec,lem:lb-unroll}).

\paragraph{Threat-based algorithms.} $\MRC_R$ belongs to the family of \emph{threat-based} online algorithms, introduced by \citet{EFKT01} for one-way trading. Such algorithms act greedily while maintaining a guarantee against the most pessimistic continuation of the input. The approach was extended to $k$-search~\citep{LPS09}, and has been used for Pareto-optimal conversion with predictions~\citep{SLHWT21}, convex function chasing with long-term constraints~\citep{LCSBHWS24}, online allocation with deadlines~\citep{LCSBHWS25}, and robust learned online policies~\citep{LYR23,LYWR23,RCMW23}.


\section{Preliminaries}\label{sec:prelim}

\paragraph{Model.} There are $n$ jobs with sizes $p_1,\dots,p_n>0$, all available at time $0$. At every time $t\ge0$, the algorithm chooses rates $r_j(t)\ge0$ with $\sum_jr_j(t)\le1$. The \emph{elapsed time} of job $j$ at time $t$ is the processing it has received so far,
\[
e_j(t)=\int_0^tr_j(u)\,du ,
\]
and job $j$ completes at the time $C_j$ when $e_j$ reaches $p_j$
\[
C_j = \inf \{ t \geq 0 \, : \, e_j(t) \geq p_j\}\;.
\]
The algorithm learns $p_j$ only at time $C_j$. The prediction is either an order $\sigma$ of the jobs, which is correct if $p_{\sigma(1)}\le\dots\le p_{\sigma(n)}$, or a vector of sizes $y$, which is correct if $y=p$. The consistency $\cons(A)$ of an algorithm $A$ is the supremum of $A/\OPT$ over all instances with a correct prediction, and its robustness $\rob(A)$ is the supremum over all instances and all predictions. Both suprema range over all numbers of jobs. For a randomized algorithm, $A$ is replaced by its expected total completion time, the instance and the prediction being fixed before the algorithm draws its random bits (oblivious adversary). 

\paragraph{The cost paid so far.} Let $N(t)$ be the number of unfinished jobs at time $t$. During a short time $dt$, the sum of completion times increases by $N(t)\,dt$, so the cost paid up to time $t$ is
\[
F(t)=\int_0^tN(u)\,du,\qquad\text{and at the end}\qquad F(\infty)=\sum_jC_j .
\]
This identity is standard in the analysis of online scheduling, for instance in amortized local competitiveness~\citep{IMP11}.

\paragraph{Hedging.} An algorithm \emph{hedges} when it processes a job while some job before it in the predicted order is unfinished. When the prediction is correct, hedging is what makes the cost exceed $\OPT$, and conversely it is what guarantees robustness when predictions are adversarial. 

\paragraph{The frozen instance.} At time $t$, the algorithm knows the sizes of the completed jobs, and every unfinished job could end immediately. The \emph{frozen instance} at time $t$ is the instance in which job $k$ has size $e_k(t)$. Its optimal total completion time~\citep{Smith56, MPT94} is given by
\[
\Phi(e)=\sum_je_j+\sum_{i<j}\min(e_i,e_j) \;.
\]
We call the terms $e_j$ the \emph{diagonal terms} of $\Phi$, and the terms $\min(e_i,e_j)$ its \emph{pair terms}.
Every instance that the algorithm cannot yet distinguish from the true one has the same sizes for the completed jobs and sizes at least $e_k(t)$ for the others. Since $\Phi$ is non-decreasing in each coordinate, its optimum is at least $\Phi(e(t))$. At the end, $\Phi(e)=\Phi(p)=\OPT$. The ratio $F(t)/\Phi(e(t))$ is the competitive ratio that the algorithm would have if all unfinished jobs ended at time $t$.


\section{Maximal Robust Commitment}\label{sec:algo}

Before formally defining our \emph{Maximal Robust Commitment} ($\MRC_R$) algorithm, we give an overview of the main ideas and invariants it maintains.

\paragraph{The robustness invariant.} $\MRC_R$ satisfies at all times the invariant
\begin{equation}\label{eq:inv}
D(t):=F(t)-R\,\Phi(e(t))\;\le\;0 .
\end{equation}
Since $F$ increases at rate $N$ whatever the algorithm does, $D$ increases at rate $N-R\,\frac{d}{dt}\Phi(e(t))$. To preserve \eqref{eq:inv} when it is tight, the algorithm must therefore make $\Phi$ grow at rate at least $N/R$.
The algorithm is based on the following observation.
\begin{quote}
\emph{If a deterministic algorithm is $R$-robust, then along every run, and in particular along its run on correct predictions, $F(t)\le R\,\Phi(e(t))$ at every time $t$} (\cref{lem:freeze-lb}).
\end{quote}
Indeed, stop the run at time $t$, and consider the adversarial instance in which every unfinished job has size $e_k(t)+\delta$ for an arbitrarily small $\delta>0$, while completed jobs keep their sizes. Up to time $t$ the algorithm observes exactly the same completions, so, being deterministic, it behaves in the same way. It has paid $F(t)$, and the adversarial instance has optimum close to $\Phi(e(t))$. \citet{WZ20} use this adversary at the first completion of the run, while we use it here at any time $t$. Therefore, every $R$-robust deterministic algorithm must satisfy the invariant \eqref{eq:inv}, even on correct predictions. $\MRC_R$, described in \Cref{alg:ft}, greedily follows the predictions as much as possible while satisfying the invariant. 

\paragraph{Committing and Round-Robin.} We call the \emph{committed job} $j$ the first unfinished job in the predicted order, and the \emph{waiting jobs} are the other unfinished jobs. In a run of $\MRC_R$, the waiting jobs always have a common elapsed time $\ell$, which we call the \emph{level}, and the committed job has $e_j\ge\ell$ (\cref{lem:structure}). The algorithm combines two moves: \emph{committing}, i.e.\ running $j$ alone, and \emph{Round-Robin} on the waiting jobs. To compute how fast each move makes $\Phi$ grow, note that a diagonal term grows at the rate at which its job is processed, and a pair term grows at the rate of the job with the smaller elapsed time, or at the smaller of the two rates if the elapsed times are equal.
\begin{itemize}
\item When $j$ runs alone at rate $1$, its diagonal term grows at rate $1$, and so does its pair term with every job $i$ with $e_i>e_j$, and no other term changes. Therefore $\Phi$ grows at the \emph{commit rate}
\[
\gamma_c=1+|\{i:e_i>e_j\}| .
\]
Completed jobs are included in this count, with elapsed time equal to their size: a completed job longer than $e_j$ is delayed by $j$ in the optimal run of the frozen instance.
\item When the $N-1$ waiting jobs share the machine equally, their diagonal terms grow at total rate $1$, their pair terms with the jobs above the level grow at total rate $|\{i\ne j:e_i>\ell\}|+1$ ($+1$ accounts for $j$), and the $\binom{N-1}2$ pair terms among them grow at total rate $\frac{N-2}2$. So $\Phi$ grows at the \emph{Round-Robin rate}
\[
\gamma_r=1+|\{i\ne j:e_i>\ell\}|+\frac N2 .
\]
In this count $j$ is treated as above the level, which is exact in every move of the algorithm, since $j$ always runs faster than the waiting jobs (\cref{lem:structure}\ref{it:st-b}).
\end{itemize}
Round-Robin is always a safe move, as $\gamma_r\ge\frac N2+1>\frac NR$ for $R\ge2$, because every unit of processing spread over the waiting jobs raises all pair terms among them. Committing can be unsafe: if no job has more elapsed time than $j$, then $\gamma_c=1$ and $D$ grows at rate $N-R$. Committing is what consistency needs, and Round-Robin is what robustness needs.

\paragraph{The algorithm.} $\MRC_R$ is fully described in \Cref{alg:ft}. We call $D$ the \emph{slack} of the algorithm. In the \emph{commit branch}, $\MRC_R$ commits, and it does so whenever this preserves the invariant, i.e.\ if $D<0$ or $\gamma_c\ge N/R$. Otherwise it is in the \emph{mixing branch}, where $D=0$ and $\gamma_c<N/R<\gamma_r$, and in this case it gives $j$ the largest rate $a$ that keeps $D=0$, namely the solution of $a\gamma_c+(1-a)\gamma_r=N/R$, and shares the rest of the machine equally among the waiting jobs. All rates are constant between consecutive \emph{events}, and there are $O(n^2)$ of them (\cref{lem:events}).

\begin{algorithm}[t]
\caption{Maximal Robust Commitment $\MRC_R$}\label{alg:ft}
\KwIn{predicted order $\sigma$, parameter $R\ge2$}
\While{some job is unfinished}{
  $N\gets$ number of unfinished jobs;\quad $j\gets$ first unfinished job in $\sigma$\;
  \lIf{$N=1$}{run $j$ until it completes, and stop}
  $\ell\gets$ the common elapsed time of the $N-1$ waiting jobs\tcp*{the level}
  $\gamma_c\gets1+|\{i:e_i>e_j\}|$;\quad $\gamma_r\gets1+|\{i\ne j:e_i>\ell\}|+\frac N2$\;
  \eIf(\tcp*[f]{committing is safe}){$D<0$ \textbf{or} $\gamma_c\ge N/R$}{
    $a\gets1$\;
  }(\tcp*[f]{commit as much as $D\le0$ allows}){
    $a\gets\dfrac{\gamma_r-N/R}{\gamma_r-\gamma_c}$\tcp*{so that $a\gamma_c+(1-a)\gamma_r=N/R$}
  }
  run $j$ at rate $a$ and each waiting job at rate $\frac{1-a}{N-1}$ \textbf{until the next event}: a job completes; $D$ increases to $0$ (from a negative value); or $e_j$ or $\ell$ reaches the size of a completed job that it did not equal at the start of the step\;
}
\end{algorithm}


\section{Overview of the analysis}\label{sec:overview}
By definition of $\MRC_R$, we have in the commit branch that either $D<0$, and the step ends as soon as $D$ increases to $0$, or $D$ grows at rate $N-R\gamma_c\le0$. In the mixing branch, $\Phi$ grows at rate exactly $a\gamma_c+(1-a)\gamma_r=N/R$, so $D$ stays at $0$. Hence $D\le0$ at all times. At the end $\Phi(e)=\OPT$, so $\MRC_R\le R\cdot\OPT$, which is the robustness bound (\cref{thm:ft-rob}). The rest of this section explains the proofs of the consistency of $\MRC_R$ and of the lower bound.

\subsection{The cost of \texorpdfstring{$\MRC_R$}{MRC} layer by layer}

Fix an instance with a correct order, and relabel the jobs so that $p_1\le\dots\le p_n$. On such an instance, we prove that $\MRC_R$ completes the jobs in this order: at every time the completed jobs are the shortest ones, $j$ is committed, and the other unfinished jobs wait at the level (\cref{lem:ft-level}).

\begin{figure}[t]
\centering
\includegraphics{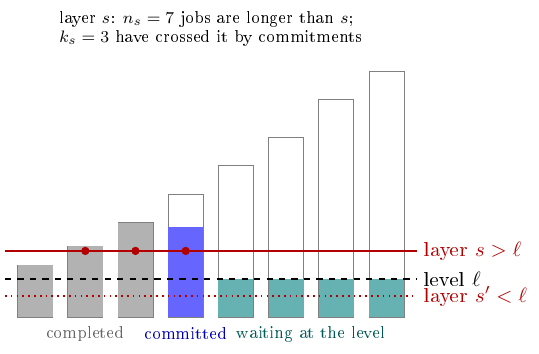}
\caption{Layers of elapsed time. Jobs are columns of height $p_j$, sorted by size and filled up to their elapsed time. A layer above the level has been crossed only by commitments; a layer below the level has been crossed by some commitments and then by one Round-Robin pass of the jobs unfinished at that moment.}
\label{fig:layers}
\end{figure}

We cut the elapsed-time axis into \emph{layers} $[s,s+ds]$, and say that a job \emph{crosses} layer $s$ when its elapsed time passes $s$ (\cref{fig:layers}). Let $n_s$ be the number of jobs of size larger than $s$. Fix a time $t$, and let $\ell$ be the level attained by the waiting jobs at this time. Moving one job through layer $s$ uses $ds$ units of machine time, during which the total completion time increases by $N\,ds$. Charging the cost in this way to layers gives
\[
F=\int_0^\infty F_s\,ds,\qquad \Phi=\int_0^\infty\Phi_s\,ds,\qquad\text{with}\qquad \Phi_s=\frac{k_s(k_s+1)}2 ,
\]
where $F_s$ is the sum of $N$ over the crossings of layer $s$, and $k_s$ is the number of jobs that have crossed layer $s$ (\cref{lem:layer-phi,lem:layer-acc}).

A layer $s$ below the level $\ell$ has been crossed by all $n_s$ jobs longer than $s$: first by commitments, one at a time, and then by the jobs still unfinished when the level reached $s$, whose number we denote by $U_s$. By analyzing the order in which the jobs cross layer $s$, we prove in \Cref{lem:layer-acc} that
\[
\Phi_s=\frac{n_s(n_s+1)}2,\qquad F_s=\frac{n_s(n_s+1)}2+\frac{U_s(U_s-1)}2 .
\]
A layer above the level has been crossed only by commitments, one at a time, as in SPT. At the end of the run, integrating these gives the cost formula (\cref{cor:cost})
\begin{equation}\label{eq:cost-main}
\MRC_R=\OPT+\int_0^\infty\frac{U_s(U_s-1)}2\,ds .
\end{equation}

\subsection{Reduction to a one-dimensional problem}

To compare layers of different instances, we measure the \emph{position} of layer $s$ by the fraction of $\OPT$ contributed by the layers below it,
\[
w(s)=\frac1{\OPT}\int_0^s\Phi_v(p)\,dv,\qquad\text{where}\qquad \Phi_v(p)=\frac{n_v(n_v+1)}2 \;.
\]
We still assume that the predicted order is correct. Let $\hat x(w(s))=1-U_s/n_s$ be the fraction of the jobs longer than $s$ that are already completed when the level reaches $s$. By \eqref{eq:cost-main}, the cost is, up to lower-order terms,
\begin{equation}\label{eq:1dim-con}
\frac{\MRC_R}\OPT\approx1+\int_0^1\big(1-\hat x(w)\big)^2dw .  
\end{equation}
For the slack $D$, when the level is at $w$, a layer below the level contributes $F_s-R\Phi_s\approx2\,\Phi_s(p)\,q(\hat x(w(s)))$, and a layer above the level, crossed by a fraction $\xi$ of the jobs longer than it, contributes $F_s-R\Phi_s\approx2\,\Phi_s(p)\,h(\xi)$, where
\[
q(\xi)=\frac{(1-\xi)^2-(R-1)}2,\qquad h(\xi)=\xi-\frac{1+R}2\xi^2 .
\]
The completed jobs are the shortest ones, so a higher layer has been crossed by a smaller fraction of its jobs, and $h$ is increasing on $[0,\tilde{\xi}]$, where $\tilde{\xi}=\frac1{1+R}$. For a function $x$ on $[0,1]$, let
\[
\Gamma_x(w)=\int_0^wq\big(x(v)\big)\,dv+(1-w)\,h\big(x(w)\big) .
\]
Then, on every instance with a correct order, as long as $\hat x(w)\le\tilde{\xi}$,
\begin{equation}\label{eq:1dim-rob}
\frac{D}{\OPT}\ \lesssim\ 2 \,\Gamma_{\hat x}(w) \;.
\end{equation}
In this sense, the consistency and robustness of $\MRC_R$ on any instance reduce respectively to the bounds \eqref{eq:1dim-con} and \eqref{eq:1dim-rob}, which depend only on the function $\hat x$.

\paragraph{Comparison with a target.} These approximate relations are made rigorous by comparing $\MRC_R$ with a fixed function $x$, a \emph{target}, rather than with its own fraction $\hat x$, which is unknown. A target prescribes the fraction of jobs that should be completed when the level is at $w$. It is \emph{admissible} if it is a non-decreasing, right-continuous function from $[0,1]$ to $[0,1]$ that stays below $\tilde{\xi}$ until some point $\tilde w\in(0,1]$ and equals $1$ afterwards, and satisfies $\Gamma_x\le0$ (with a boundary condition at $\tilde w$, see \eqref{eq:target}). \Cref{thm:budget} shows that, on every instance with a correct predicted order,
\begin{equation}\label{eq:target-bound}
\MRC_R\ \le\ \Big(1+\int_0^1\big(1-x(w)\big)^2dw\Big)\OPT\qquad\text{for every admissible target }x .
\end{equation}

The proof defines the \emph{budget} $T_s=\Phi_s(p)\,(1-x(w(s)))^2$ of layer $s$ and shows that $\frac{U_s(U_s-1)}2\le T_s$ in almost every layer. Then, \eqref{eq:target-bound} follows from \eqref{eq:cost-main} and $\Phi_s(p)\,ds=\OPT\,dw$. 
The argument is by contraposition: $\MRC_R$ raises its level only when committing is unsafe, so it is never over budget. Consider the first time $t$ at which $\MRC_R$ raises its level while $\frac{N(t)(N(t)-1)}2\ge T_{\ell(t)}$, i.e.\ while its number of unfinished jobs is over the budget of the layer at the level. All layers below the level were passed within budget, so they contribute to $D$ at most what the target allows; and since the number of unfinished jobs is over budget, few jobs have crossed the layers above the level. Summing, committing at that time would keep $D / \OPT$ below $2 \, \Gamma_x \leq 0$. But $\MRC_R$ raises its level only when committing alone would make $D$ positive, which is a contradiction. 

\subsection{The optimal target}

By \eqref{eq:target-bound}, it remains to find an admissible target with a small value of $\int_0^1(1-x)^2$. To make this integral small, $x$ should be as large as possible, subject to $\Gamma_x\le0$. Since $\Gamma_x(0)=h(x(0))$ and $h>0$ on $(0,\tilde{\xi}]$, the target starts at $x(0)=0$. For a differentiable target,
\begin{align*}
\Gamma_x'(w)
&=q\big(x(w)\big)-h\big(x(w)\big)+(1-w)\,h'\big(x(w)\big)\,x'(w) \\
&=-\frac{R+2}2\big(1-x\big)\big(x+\beta\big)+(1-w)\big(1-(1+R)x\big)x',
\end{align*}
where $x=x(w)$ and $\beta=\frac{R-2}{R+2}$. The first term, due to the level rising, is negative. The second, due to committing, is positive as long as $x<\tilde{\xi}$. Keeping $\Gamma_x=0$, that is, committing as fast as the constraint allows, gives the differential equation
\[
\frac{dx}{dw}=\frac{(R+2)(1-x)(x+\beta)}{2(1-w)\,\big(1-(1+R)x\big)},\qquad x(0)=0 \;,
\]
until $x$ reaches $\tilde{\xi}$. From then on, committing decreases $\Gamma_x$, and the target can jump to $1$. The solution of this equation, the \emph{profile} $x^*$, is given implicitly by
\[
1-w=\Big[(1-x^*)\Big(1+\frac{x^*}\beta\Big)^{\frac R{R+2}}\Big]^{-1}\qquad\text{for }x^*<\tilde{\xi} .
\]
This derivation only serves to find $x^*$. \Cref{lem:profile} verifies directly that $x^*$ is admissible and computes
\[
\int_0^1\big(1-x^*(w)\big)^2dw=c(R)-1 ,
\]
which, with \eqref{eq:target-bound}, proves the consistency bound of \cref{thm:ub}.

\subsection{The lower bound}

Consider $n$ jobs of size $1$ with correct predicted sizes, and a deterministic $R$-robust algorithm. By the robustness invariant, $F\le R\Phi$ must hold at each completion time $C_1\le\dots\le C_n$ of its run. Let $E_k$ be the hedging done before the $k$-th completion, that is, the processing received by the jobs that are still unfinished at that time, and $S_k=E_1+\dots+E_k$ the cumulated hedging. Every unit of $E_k$ delays the completions, so $\sum_kC_k-\OPT\ge S_{n-1}$. Bounding each pairwise minimum in $\Phi$ by an average, the constraint at the $k$-th completion becomes
\[
d_k\,E_k\ \ge\ a_k+S_{k-1}
\]
for explicit $a_k$ and $d_k$: the hedging forced at a completion grows with the hedging already done. This recursion is used until the completed fraction reaches $\tilde{\xi}=\frac1{R+1}$, and the constraint at that point bounds the hedging of the remaining jobs. Unrolling the recursion and letting $n\to\infty$ shows that the relative excess cost $(\sum_kC_k-\OPT)/\OPT$ is at least an explicit integral, which equals $c(R)-1$. Hence the consistency of the algorithm is at least $c(R)$ (\cref{thm:lb}).

\begin{remark}
Near consistency $1$, the smallest robustness $R^*(1+\eps)$ of a deterministic algorithm with consistency $1+\eps$ satisfies $R^*(1+\eps)=\eps^{-1/2}+\frac16+O(\sqrt\eps)$ (\cref{cor:frontier}). At consistency $1+\eps$, previous algorithms achieve robustness $1/\eps+2$ (time sharing~\citep{PSK18}) and $1/\eps+1$~\citep{BCLS25}. The frontier is of order $\eps^{-1/2}$, like the lower bound $1/\sqrt{2\eps}$ of~\citep{WZ20}, which is asymptotically smaller than $R^*$ by a factor $\sqrt2$. 
\end{remark}

\subsection{Randomization}\label{sec:rand-overview}

With predicted sizes $y$, sorted as $y_1\le\dots\le y_n$, let $n_s=|\{j:y_j>s\}|$ and $w_y(s)=\frac1{\Phi(y)}\int_0^s\tfrac{n_u(n_u+1)}2\,du$, the fraction of the predicted optimum $\Phi(y)$ paid by the layers below $s$: this is the position $w(s)$ of \cref{sec:overview}, computed from $y$. In \cref{alg:rt}, a \emph{mismatch} is a job that completes with elapsed time different from its predicted size, or that reaches it without completing.

\begin{algorithm}[t]
\caption{Randomized two-stage algorithm $\RT_\rho$}\label{alg:rt}
\KwIn{predicted sizes $y$, parameter $\rho\ge\rho_0$}
draw $V$ uniformly in $(0,1)$, and let $B\gets V^2/(V^2+3\rho V+\rho^2)$ and $L\gets w_y^{-1}(B)$\;
run Round-Robin until the common elapsed time of the unfinished jobs reaches $L$\;
run the unfinished jobs alone, one at a time, in increasing order of $y_j$\;
at the first mismatch, if any, run Round-Robin on the unfinished jobs of smallest elapsed time until all jobs complete\;
\end{algorithm}

\emph{Consistency.} With correct predictions there is no mismatch, the layers below $L$ are crossed in Round-Robin and those above by commitments, so the cost is $\Phi(y)+\int_0^Ln_s(n_s-1)/2\,ds$, which has the form of \eqref{eq:cost-main} with $U_s=n_s$ for $s<L$, and is at most $(1+B)\,\Phi(y)$ by definition of $L$. Hence the consistency is at most $1+\E[B]=1+\eps(\rho)$.

\emph{Robustness.} For a fixed instance and prediction, let $d$ be the smallest elapsed time at which a mismatch can occur. If $L\ge d$, the run is Round-Robin, with cost at most $2\,\OPT$. If $L<d$, comparing the run with Round-Robin pair of jobs by pair of jobs, and bounding each layer, shows that the excess over Round-Robin is at most $\max\{\hat g(w_y(d)),\rho_0\}\,\OPT$ for an explicit decreasing function $\hat g$, independent of $L$. The law of $B$ is chosen so that $\Prob(B<w)\max\{\hat g(w),\rho_0\}\le\rho$ for every $w$, so the expected cost is at most $(2+\rho)\OPT$. The scale of $L$ and the detection of errors both need predicted sizes.


\section{Experiments}\label{sec:exp}

In this section, we empirically compare $\MRC_R$ with previous algorithms on random instances, following the setups of~\citep{PSK18,LM22,BP24}. Job sizes are drawn independently from the exponential distribution with mean $1$, from the Pareto distribution with scale $1$ and shape $1.1$, or from the Weibull distribution with scale $2$ and shape $0.5$~\citep{LM22}. The predicted size of job $i$ is $y_i=\max\{0,\,p_i+\eps_i\}$, where the noise terms $\eps_i$ are independent normal variables with mean $0$ and standard deviation $\tau$, and the predicted order sorts the jobs by $y_i$. Each point averages $2000$ independent instances.

\begin{figure}[h]
\centering
\includegraphics[width=\textwidth]{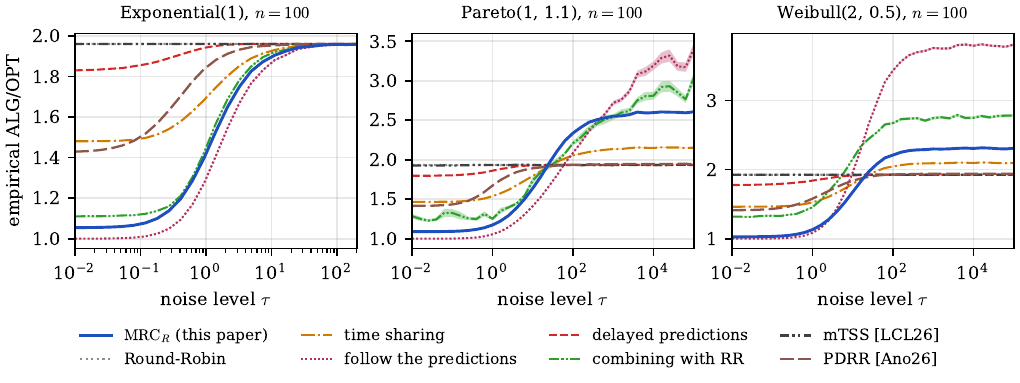}
\caption{Empirical ratio to $\OPT$ as a function of the noise level $\tau$ (logarithmic scale), for $R=3$ and $n=100$, with $2000$ independent instances per point. The bands ($\pm1$ standard error) are mostly thinner than the lines.}
\label{fig:exp}
\end{figure}

We compare $\MRC_R$ with four other $R$-robust algorithms, time sharing~\citep{PSK18}, the delayed-predictions algorithm of~\citep{BCLS25}, and the algorithms of the concurrent works~\citep{LCL26} (mTSS) and~\citep{Ano26} (PDRR), and with three algorithms that do not depend on $R$: Round-Robin (prediction agnostic), following the predicted order (not robust), and the randomized combining algorithm of~\citep{BCLS25}, which selects Round-Robin or the predicted order based on a random sample of jobs. We use $R=3$ and $n=100$ jobs (details in \cref{app:exp}); \cref{fig:exp} shows the results.

With accurate predictions, $\MRC_3$ is close to following the predictions on all three distributions, and far below time sharing. Delayed predictions, with the parameters that give it the best consistency at robustness $3$, approaches Round-Robin as soon as $\tau>0$. For large $\tau$, the predicted order is essentially random. On exponential sizes, all algorithms then approach Round-Robin. On Pareto and Weibull sizes, following the predictions and the combining algorithm become worse than Round-Robin, while the $3$-robust algorithms level off below their guarantee. There, $\MRC_3$ ends above time sharing, because it uses more of the slack allowed by the robustness constraint to follow the predictions, but still falls below its nominal robustness. The algorithms of the concurrent works are far from following the predictions when they are accurate, and quickly behave as Round-Robin when they are not. The algorithm mTSS of~\citep{LCL26} has ratio $1+\lambda\approx1.19$ on every instance with exact predictions, since it spends its whole consistency budget by design, and it coincides with Round-Robin for every $\tau>0$: it returns to Round-Robin at the first prediction error, and with continuous noise this happens at the first completion. In practice, predicted sizes are never exactly correct, so mTSS cannot benefit from them without the tolerance mechanism of~\citep{LCL26}, which weakens it in practice. The algorithm PDRR of~\citep{Ano26} is conservative: its ratio is about $1.42$ with accurate predictions, comparable to time sharing, and it approaches Round-Robin already for moderate noise, for instance $1.88$ at $\tau\approx1.4$ on exponential sizes, against $1.53$ for $\MRC_3$. Its parameter is dictated by its robustness guarantee $\min\{2\eps^{-1/2},2+\eps^{-1/2}\}=3$, while the worst instance known for it~\citep{Ano26} has ratio only about $2.33$ at this parameter, so its guarantee may be loose in this regime. In comparison, $\MRC_3$ has ratio between $1.03$ and $1.09$ with accurate predictions, and remains below both algorithms on exponential sizes for every $\tau$, and on Pareto and Weibull sizes for $\tau$ up to about $25$. Only for larger noise on these heavy-tailed sizes, where the predicted order carries little information, do mTSS and PDRR end below $\MRC_3$, as they stay close to Round-Robin.

\subsection{Setup}\label{app:exp}

\paragraph{Instances.} Each point averages $2000$ independent instances (sizes and predictions); every algorithm, including the randomized combining algorithm of~\citep{BCLS25}, is run once on each instance.

\paragraph{Algorithms.} For a robustness parameter $R>2$, the $R$-robust algorithms are time sharing~\citep{PSK18}, which runs the job with the smallest predicted size at rate $\lambda=1-\frac2R$ and Round-Robin at rate $1-\lambda$; the delayed-predictions algorithm of~\citep{BCLS25}, which runs Shortest Elapsed Time First and, when job $j$ reaches elapsed time $\alpha y_j$, runs it alone until its elapsed time is $y_j/\rho$; mTSS~\citep[Algorithm~1]{LCL26}; PDRR~\citep{Ano26}; and $\MRC_R$. Delayed predictions is $(1+\alpha)$-consistent and $(1+\frac1{\rho\alpha})$-robust for $\alpha,\rho\in(0,1]$~\citep{BCLS25}; at robustness $R$, the best consistency is obtained with $\rho=1$ and $\alpha=\frac1{R-1}$, which we use. For mTSS, we use the smallest consistency parameter $\lambda$ whose robustness guarantee $\Psi_n(\lambda)$~\citep[Theorem~4.4]{LCL26} is at most $R$, which gives $\lambda\approx0.188$ for $n=100$ and $R=3$. This guarantee is attained on instances with equal predicted sizes, so mTSS is not Pareto-optimal: for instance, with $n=100$ and consistency $c(3)\approx1.156$, its robustness is about $3.2$, against $3$ for $\MRC_3$. In mTSS, a prediction error is detected when a job completes with elapsed time different from its predicted size, or reaches its predicted size without completing. For PDRR, we use the smallest drift parameter $\eps$ whose robustness guarantee $\min\{2\eps^{-1/2},2+\eps^{-1/2}\}$ is at most $R$, that is, $\eps=\min\{\frac4{R^2},\frac1{(R-2)^2}\}=\frac49$ for $R=3$. The robustness of PDRR is at least $\eps^{-1/2}+\frac12+\frac{\sqrt\eps}2$~\citep{Ano26}, which exceeds $R^*(1+\eps)$ for every $\eps\in(0,1)$, so PDRR is not Pareto-optimal either. The other algorithms do not depend on $R$: Round-Robin; following the predictions, which runs the jobs one after the other in the predicted order and is not robust; and the randomized combining algorithm of~\citep{BCLS25}, which runs $\lceil n^{2/3}(\ln2)^{1/3}/8\rceil$ random pairs of jobs to completion, and then follows Round-Robin or the predicted order, whichever has the smaller pairwise delay on the sample.

\section{Discussion}\label{sec:discussion}

In this work, we showed that, with predicted job sizes as well as with a predicted order, the best consistency of a deterministic $R$-robust algorithm is $c(R)$. The Pareto-optimal trade-off is therefore the curve $\{(c(R),R):R>2\}$, and it is attained by $\MRC_R$, which only uses the predicted order. Furthermore, we proved that, with size predictions, randomized algorithms can achieve strictly better tradeoffs than the Pareto-optimal frontier for deterministic algorithms.  We conclude with some open questions.

\paragraph{Smoothness.} Consistency and robustness describe the two extremes of correct and adversarial predictions. In between, the cost should ideally degrade gracefully with the prediction error, for instance as $c(R) \OPT$ plus a multiple of the inversion error $\eta=\sum(p_i-p_k)$ over the pairs predicted in the wrong order~\citep{LM22}. This is the smoothness criterion for learning-augmented algorithms. An algorithm is called \emph{brittle} if an arbitrarily small prediction error can raise its cost from the consistency bound to the robustness bound, and Pareto-optimal algorithms can be brittle~\citep{EADL24,BP25}. Whether $\MRC_R$ is smooth is open. If it is not, is there a Pareto-optimal algorithm that is smooth? And what is the best smoothness guarantee subject to Pareto-optimality of the consistency and robustness?

\paragraph{Other scheduling models.} The robustness invariant extends to weighted completion times, release dates and parallel machines: the frozen instance is still well defined, and $F\le R\Phi$ is still necessary and sufficient for $R$-robustness. Its optimum $\Phi$ is computed by Smith's ratio rule for weighted jobs~\citep{Smith56}, by the shortest remaining processing time rule with release dates~\citep{Sch68}, and by SPT on identical parallel machines~\citep{McN59}. We expect our greedy algorithm, and possibly its layer-based analysis, to extend to these variants.

\paragraph{Randomized algorithms.} With predicted sizes, randomization helps (\cref{thm:rand}). The exact randomized frontier, whether consistency $1+o(1)$ and robustness $2+o(1)$ can be achieved simultaneously, and whether randomization helps with a predicted order only, are open.

\addcontentsline{toc}{section}{References}
\bibliographystyle{alphanames}
\bibliography{references}

\newpage
\appendix
\crefalias{section}{appendix}
\crefalias{subsection}{appendix}
\part*{Appendix: complete proofs}
\addcontentsline{toc}{part}{Appendix: complete proofs}

The appendix proves the upper bound (\cref{thm:ub}) and the lower bound (\cref{thm:lbm}) in separate sections. \Cref{app:prelim} collects the model, the notation (\cref{tab:terms}) and the layer representation of $\Phi$. \Cref{app:run} proves the properties of the run of $\MRC_R$ used by both parts of the upper bound; \cref{app:ft} proves robustness (\cref{thm:ft-rob}) and \cref{app:cons} consistency (\cref{thm:ft-cons}). \Cref{app:lbsec} proves the lower bound (\cref{thm:lb}). Finally, \cref{app:frontier} derives the behavior of the trade-off near consistency $1$ (\cref{cor:frontier}), \cref{sec:exp} reports the experiments, and \cref{app:rand} proves \cref{thm:rand} on randomized algorithms.

\section{Preliminaries}\label{app:prelim}

\paragraph{Model.} An instance has $n$ jobs with sizes $p\in(0,\infty)^n$, and $\OPT(p)$ is its optimal total completion time, attained by SPT. The prediction is a permutation $\sigma$ of the jobs, a predicted order; it is correct if $p_{\sigma(1)}\le\dots\le p_{\sigma(n)}$. (Predicted sizes are needed only for the lower bound and for the randomized algorithm of \cref{app:rand}, and are introduced in \cref{app:lbsec}.) A deterministic algorithm knows $n$ and $\sigma$, and at any time knows which jobs have completed and when. Formally, it assigns to every completion history $H=((j_1,C_1),\dots,(j_k,C_k))$, with $C_1\le\dots\le C_k$ (simultaneous completions listed by index; $C_0:=0$ for the empty history), a measurable rate function on $[C_k,\infty)$, with non-negative rates summing to at most $1$ and supported on the jobs not in $H$. The run is defined by induction over completions: after the $k$-th completion, the algorithm follows the rate function of the current history until the next completion. Writing $A(p,\sigma)$ for the total completion time of the run of $A$ on sizes $p$ with prediction $\sigma$,
\[
\cons(A)=\sup_{n\ge1}\ \sup_{\substack{p\in(0,\infty)^n,\ \sigma\\ \sigma\text{ correct for }p}}\frac{A(p,\sigma)}{\OPT(p)},\qquad
\rob(A)=\sup_{n\ge1}\ \sup_{p\in(0,\infty)^n,\ \sigma}\frac{A(p,\sigma)}{\OPT(p)} .
\]
A randomized algorithm draws its random bits from a fixed distribution, independently of $(p,\sigma)$, and then runs a deterministic algorithm depending on these bits. Its consistency and robustness are defined by the same formulas, with $A(p,\sigma)$ replaced by its expectation over the random bits. This is the oblivious adversary, which fixes the instance and the prediction in advance.

\paragraph{Notation.} For $e\in\mathbb R_{\ge0}^n$ let
\[
\Phi(e)=\sum_je_j+\sum_{i<j}\min(e_i,e_j).
\]
Smith's rule~\citep{Smith56} gives $\OPT(p)=\Phi(p)$ (see also~\citep{MPT94}). A schedule assigns to every job a measurable rate $r_j(t)\ge0$, with $\sum_jr_j(t)\le1$. It is \emph{non-idling} if $\sum_jr_j(t)=1$ for almost every $t$ at which some job is unfinished. We write
\[
e_j(t)=\int_0^tr_j(u)\,du,\qquad C_j=\inf\{t:e_j(t)\ge p_j\},\qquad N(t)=|\{j:C_j>t\}|
\]
for the elapsed time of job $j$ (the processing it has received by time $t$), its completion time, and the number of unfinished jobs. The cost so far is
\[
F(t)=\int_0^tN(u)\,du .
\]
Since $N(u)=\sum_j\one[C_j>u]$, we have $F(\infty)=\sum_jC_j$: the total completion time increases at rate $N(t)$. Each run of a deterministic algorithm is a schedule in this sense; on $[0,t]$ it depends only on the prediction and on the completions that occurred in $[0,t]$.

\Cref{tab:terms} collects the notation used in \cref{app:run,app:ft,app:cons,app:lbsec,app:frontier}. \Cref{app:rand} introduces its own notation.

\begin{table}[h]
\centering
\small
\renewcommand{\arraystretch}{1.15}
\begin{tabular}{@{}p{0.25\linewidth}p{0.71\linewidth}@{}}
\toprule
notation & meaning\\\midrule
$e_j(t)$ & elapsed time of job $j$: processing received up to time $t$\\
$N(t)$ & number of unfinished jobs at time $t$\\
$F(t)$ & cost so far, $\int_0^tN(u)\,du$; at the end, $F=\sum_jC_j$\\
$\Phi(e)$ & optimum of the frozen instance, in which job $k$ has size $e_k$\\
$D$ & slack $F-R\Phi$; $\MRC_R$ keeps $D\le0$\\
committed job $j$ & first unfinished job in the predicted order $\sigma$\\
waiting jobs & unfinished jobs other than $j$\\
level $\ell$ & common elapsed time of the waiting jobs\\
committing & running $j$ alone\\
Round-Robin & sharing the machine equally among the waiting jobs\\
$\gamma_c$, $\gamma_r$ & commit rate and Round-Robin rate of $\Phi$ (\cref{lem:phirate})\\
layer $s$ & a value $s\ge0$ of elapsed time; a job crosses it when its elapsed time passes $s$\\
$n_s$, $k_s$ & number of jobs with size $>s$, resp.\ with elapsed time $>s$\\
$U_s$ & number of unfinished jobs when the level first reaches $s$\\
$w(s)$ & fraction of $\OPT$ contributed by the layers below $s$\\
target $x$, $\Gamma_x$ & a function on $[0,1]$ and the quantity bounded in \eqref{eq:target}\\
$T_s$ & budget $\Phi_s(p)\,(1-x(w(s)))^2$ of layer $s$\\
$x^*$ & the profile, the target used for the consistency bound (\cref{app:profile})\\
\bottomrule
\end{tabular}
\caption{Notation used in the appendix.}
\label{tab:terms}
\end{table}

\begin{lemma}[Layer representation]\label{lem:layer-phi}
For $e\in\mathbb R_{\ge0}^n$ and $s\ge0$, let $k_s(e)=|\{j:e_j>s\}|$ be the number of jobs whose elapsed time exceeds $s$, and $\Phi_s(e)=\frac{k_s(e)(k_s(e)+1)}2$. Then
\[
\Phi(e)=\int_0^\infty\Phi_s(e)\,ds .
\]
\end{lemma}
\begin{proof}
Every real $a\ge0$ satisfies $a=\int_0^\infty\one[a>s]\,ds$. Applied to $a=e_j$ and to $a=\min(e_i,e_j)$, with $\one[\min(e_i,e_j)>s]=\one[e_i>s]\,\one[e_j>s]$, this gives
\[
e_j=\int_0^\infty\one[e_j>s]\,ds,\qquad
\min(e_i,e_j)=\int_0^\infty\one[e_i>s]\,\one[e_j>s]\,ds .
\]
Summing over $j$ and over pairs $i<j$, and exchanging the finite sums with the integral,
\[
\Phi(e)=\int_0^\infty\Big(\sum_j\one[e_j>s]+\sum_{i<j}\one[e_i>s]\,\one[e_j>s]\Big)ds .
\]
Fix $s$ and write $k_s=k_s(e)$. The first sum counts the jobs with $e_j>s$, so it equals $k_s$. The second sum counts the pairs of jobs that both have elapsed time $>s$, so it equals $\binom{k_s}2$. Hence the integrand is $k_s+\frac{k_s(k_s-1)}2=\frac{k_s(k_s+1)}2=\Phi_s(e)$.
\end{proof}


\section{The run of $\MRC_R$}\label{app:run}

This section collects the properties of the run of $\MRC_R$ that the robustness and consistency proofs use. Throughout, $R\ge2$, as in \cref{alg:ft}, and the predicted order $\sigma$ is arbitrary (not necessarily correct). \Cref{lem:structure} shows that the waiting jobs always share a common level, with the committed job at or above it; \cref{lem:phirate} computes how fast $\Phi$ grows; \cref{lem:events} shows that the run is well defined, with $D'=0$ in the mixing branch.

\paragraph{Notation.} At time $t$, the \emph{committed job} $j$ is the first unfinished job in $\sigma$ (throughout, $j$ denotes this job), $N$ is the number of unfinished jobs, and the \emph{waiting jobs} are the $N-1$ unfinished jobs other than $j$. A job $i$ is completed if and only if $e_i=p_i$; it receives no processing afterwards. \Cref{alg:ft} uses the common elapsed time $\ell$ of the waiting jobs, the \emph{level} (\cref{lem:structure} shows that they always have one), and the rates
\[
\gamma_c=1+|\{i:e_i>e_j\}|,\qquad
\gamma_r=1+|\{i\ne j:e_i>\ell\}|+\frac N2 ,
\]
where the counts include completed jobs. We write $D(t)=F(t)-R\,\Phi(e(t))$, the \emph{slack}; the invariant \eqref{eq:inv} is $D\le0$. Since $F$ increases at rate $N$, $D$ changes at rate $N-R\,\frac{d}{dt}\Phi(e(t))$.

\paragraph{The run, interval by interval.} Let $t_0=0$. Given the state at an event time $t_k$ with $N\ge2$, the algorithm computes its branch and $a$ ($a=1$ in the commit branch, $a=\frac{\gamma_r-N/R}{\gamma_r-\gamma_c}$ in the mixing branch), and runs $j$ at rate $a$ and each waiting job at rate $b=\frac{1-a}{N-1}$ until $t_{k+1}$, the first time $t>t_k$ at which one of the following \emph{events} occurs:
\begin{itemize}
\item a job completes;
\item $D$ reaches $0$ (only if $D(t_k)<0$);
\item $e_j$ reaches the size of a completed job $i$ with $p_i\ne e_j(t_k)$;
\item $\ell$ reaches the size of a completed job $i$ with $p_i\ne\ell(t_k)$.
\end{itemize}
If $N=1$ at $t_k$, the last job runs alone until it completes, and there are no further events (this final completion is not counted as an event). Let
\[
t_\infty=\sup_kt_k .
\]

\begin{lemma}[Structure of the run]\label{lem:structure}
For every instance and every $\sigma$, at every time $t<t_\infty$ with $N(t)\ge2$:
\begin{enumerate}[label=(\alph*)]
\item\label{it:st-a} the waiting jobs all have the same elapsed time $\ell$, and $e_j\ge\ell$; while a given job is the committed job, $e_j-\ell$ is strictly increasing. Consequently, the jobs counted in $\gamma_c$ and in $\gamma_r$ are completed jobs;
\item\label{it:st-b} in the mixing branch, $\gamma_c<\frac NR<\gamma_r$, so $a$ is well defined and $\frac1N<a<1$, and $b<\frac1N<a$.
\end{enumerate}
Moreover each interval $[t_k,t_{k+1})$ is non-empty and $t_{k+1}<\infty$.
\end{lemma}
\begin{proof}
\noindent\textbf{Step 1: (b) in a state as in (a).} Let $C=|\{i\ne j:e_i>\ell\}|\ge0$, so $\gamma_r=1+C+\frac N2\ge\frac N2+1$. In the mixing branch $\gamma_c<\frac NR$ (the branch condition), and
\[
\frac NR\le\frac N2<\frac N2+1\le\gamma_r
\]
since $R\ge2$. So the numerator and the denominator of $a$ are positive, and $a<1$. For $a>\frac1N$: since $R\ge2$ and $\gamma_c\ge1$,
\[
a\ \ge\ \frac{\gamma_r-N/2}{\gamma_r-1},\qquad\text{and}\qquad \frac{\gamma_r-N/2}{\gamma_r-1}>\frac1N\iff(N-1)\gamma_r>\frac{N^2}2-1 ,
\]
which holds because $(N-1)\gamma_r\ge(N-1)\big(\frac N2+1\big)=\frac{N^2}2+\frac N2-1$. Finally $b=\frac{1-a}{N-1}<\frac{1-1/N}{N-1}=\frac1N$.

\noindent\textbf{Step 2: induction over the intervals.} At time $0$ all elapsed times are $0$, so (a) holds. Suppose (a) holds at $t_k$, with $N\ge2$. Each event says that a continuous function of $t$ reaches a value that it does not have at $t_k$, so $t_{k+1}>t_k$; and $t_{k+1}<\infty$, since $j$ runs at rate $a>0$ (by Step~1 in the mixing branch) and completes at the latest at time $t_k+(p_j-e_j(t_k))/a$. On $[t_k,t_{k+1}]$ the waiting jobs run at the common rate $b$, so they keep a common elapsed time $\ell$; and $b<a$ (in the commit branch $b=0<1=a$; in the mixing branch by Step~1), so $e_j-\ell$ is strictly increasing. At $t_{k+1}$, if $j$ completes, the new committed job is the first unfinished job in $\sigma$, which is a waiting job, at elapsed time $\ell$; if waiting jobs complete, the other waiting jobs stay at $\ell$. In both cases (a) holds at $t_{k+1}$ (if $N\ge2$ there). By induction, (a) holds as long as the run is defined, and (b) holds by Step~1.

\emph{The consequence in (a).} The waiting jobs have elapsed time $\ell\le e_j$, so they are counted neither in $\{i:e_i>e_j\}$ nor in $\{i\ne j:e_i>\ell\}$, and $j$ is not counted in either set. So only completed jobs are counted.
\end{proof}

\begin{lemma}[Growth rate of $\Phi$]\label{lem:phirate}
Consider the state of the run at a time $t<t_\infty$ with $N(t)\ge2$. Suppose that, from this state, $j$ runs at rate $a$ and each waiting job at rate $b=\frac{1-a}{N-1}$, where $\frac1N\le a\le1$. Then $\Phi(e(t))$ has right derivative $a\gamma_c+(1-a)\gamma_r$. In particular, committing ($a=1$) makes $\Phi$ grow at rate $\gamma_c$.
\end{lemma}
\begin{proof}
By \cref{lem:structure}\ref{it:st-a}, the waiting jobs have a common elapsed time $\ell$, $e_j\ge\ell$, and the $C=|\{i\ne j:e_i>\ell\}|$ jobs counted in $\gamma_r$ are completed. $\Phi$ is a sum of diagonal terms $e_k$ and pair terms $\min(e_i,e_k)$. Under constant rates $r$, the right derivative of $e_k$ is $r_k$; that of $\min(e_i,e_k)$ is $r_i$ if $e_i<e_k$ (the minimum stays $e_i$ for a short time), and $\min(r_i,r_k)$ if $e_i=e_k$. Here $r_j=a$, the waiting jobs have rate $b$, and completed jobs have rate $0$; note that $b\le\frac{1-1/N}{N-1}=\frac1N\le a$. Term by term:
\begin{itemize}
\item diagonal terms: $a+(N-1)b=a+(1-a)=1$;
\item pairs $\{j,i\}$ with $i$ completed: rate $a$ if $e_i>e_j$, and $0$ if $e_i\le e_j$ (then $i$ has the smaller or equal elapsed time and rate $0$). These jobs $i$ are exactly the $\gamma_c-1$ jobs counted in $\gamma_c$, so the total is $a(\gamma_c-1)$;
\item pairs $\{g,i\}$ with $g$ waiting and $i$ completed: rate $b$ if $e_i>\ell$, and $0$ otherwise; in total $(N-1)\,C\,b=(1-a)C$;
\item the $\frac{(N-1)(N-2)}2$ pairs of two waiting jobs: equal elapsed times and equal rates, so rate $b$ each; in total $\frac{(N-1)(N-2)}2b=(1-a)\frac{N-2}2$;
\item pairs $\{j,g\}$ with $g$ waiting: rate $b$ if $e_j>\ell$, and $\min(a,b)=b$ if $e_j=\ell$ (the instant $j$ becomes the committed job); in total $(N-1)b=1-a$;
\item pairs of two completed jobs: rate $0$.
\end{itemize}
Adding up, and writing the first term $1$ as $a+(1-a)$,
\begin{align*}
\frac{d^+}{dt}\Phi(e(t))&=a+a(\gamma_c-1)+(1-a)\Big(1+C+\frac{N-2}2+1\Big)\\
&=a\gamma_c+(1-a)\Big(1+C+\frac N2\Big)\\
&=a\gamma_c+(1-a)\gamma_r .
\qedhere
\end{align*}
\end{proof}

\begin{lemma}[Events]\label{lem:events}
For every instance and every $\sigma$, the run of $\MRC_R$ has $O(n^2)$ events and is defined at all times, so $t_\infty$ is the last event time; in particular \cref{lem:structure,lem:phirate} apply at every time with $N(t)\ge2$. Moreover:
\begin{enumerate}[label=(\alph*)]
\item\label{it:ev-a} between consecutive events, $N$, $\gamma_c$ and $\gamma_r$ are constant, and $\Phi(e(t))$ is affine, with slope $\gamma_c$ in the commit branch and $\frac NR$ in the mixing branch. Hence $D'=N-R\gamma_c$ in the commit branch and $D'=0$ in the mixing branch;
\item\label{it:ev-b} at the last event time at most one job is unfinished; if one is, it runs alone until it completes.
\end{enumerate}
\end{lemma}
\begin{proof}
\noindent\textbf{Step 1: between consecutive events.} On $[t_k,t_{k+1})$ the rates are constant, and $N$ is constant since completions are events. The jobs counted in $\gamma_c$ and in $\gamma_r$ are completed jobs (\cref{lem:structure}\ref{it:st-a}), with fixed elapsed times equal to their sizes. Inside the interval, $e_j$ and $\ell$ do not reach any of these sizes (this would be an event); a completed job with $p_i=e_j(t_k)$ is not counted in $\gamma_c$ at $t_k$, nor afterwards, as $e_j$ increases, and similarly for $\ell$ and $\gamma_r$. So $\gamma_c$ and $\gamma_r$ are constant on $[t_k,t_{k+1})$. Since $\frac1N\le a\le1$ ($a=1$ in the commit branch, and \cref{lem:structure}\ref{it:st-b} in the mixing branch), \cref{lem:phirate} gives the right derivative $a\gamma_c+(1-a)\gamma_r$ of $\Phi(e(t))$ at every point of the interval, a constant. A continuous function with constant right derivative on an interval is affine there, so $\Phi(e(t))$ is affine with slope $\gamma_c$ in the commit branch ($a=1$) and
\[
a\gamma_c+(1-a)\gamma_r=\gamma_r-a(\gamma_r-\gamma_c)=\frac NR
\]
in the mixing branch, by the choice of $a$. Since $F'=N$, this gives the values of $D'$.

\noindent\textbf{Step 2: counting events.} Call a \emph{commit episode} the time during which a given job $j$ is the committed job; there are at most $n$ episodes. Fix one. During it, $e_j$ is strictly increasing ($a>0$) and $\ell$ is non-decreasing, so each of them reaches each of the at most $n$ sizes of completed jobs at most once; together with the at most $n$ completions in total, this gives $O(n)$ events of these types, and between two consecutive ones $N$ and $\gamma_c$ are constant. In between, there is at most one event ``$D$ reaches $0$'': it occurs in the commit branch with $D<0$ and $D'=N-R\gamma_c>0$, so afterwards $D=0$ and $\gamma_c<N/R$, the algorithm mixes, and $D'=0$ keeps $D=0$, until $N$ or $\gamma_c$ changes. Hence each episode has $O(n)$ events, and the run has $O(n^2)$ events.

\noindent\textbf{Step 3: after the last event.} At the last event time $N\le1$ (the completion of a job running alone is not an event): otherwise the construction would continue, since $t_{k+1}$ exists and is finite (\cref{lem:structure}). If $N=1$, the last job runs alone until it completes. So the run is defined at all times, and \cref{lem:structure} holds at all times.
\end{proof}


\section{Robustness of $\MRC_R$}\label{app:ft}

This section proves that $\MRC_R$ is $R$-robust: for every instance and every predicted order, $\MRC_R\le R\,\OPT$. In the mixing branch $\Phi$ grows at exactly the required rate $N/R$, so $D$ stays constant, and the commit branch is used only when $D<0$ or when committing alone keeps $D'\le0$.

\begin{theorem}[Robustness for every prediction]\label{thm:ft-rob}
Let $R\ge2$. For every instance and every permutation $\sigma$, $\MRC_R$ maintains $F(t)\le R\,\Phi(e(t))$ for all $t$. In particular $\rob(\MRC_R)\le R$.
\end{theorem}
\begin{proof}
$D(t)=F(t)-R\Phi(e(t))$ is continuous, and $D(0)=0$. By \cref{lem:events}, there are finitely many events; we show by induction that $D(t_k)\le0$ implies $D\le0$ on $[t_k,t_{k+1}]$. On this interval $D$ is affine (\cref{lem:events}\ref{it:ev-a}).
\begin{itemize}
\item \emph{$D(t_k)<0$.} The algorithm commits. If $D'\le0$, $D$ stays negative. If $D'>0$, the interval ends at the latest when $D$ reaches $0$, since this is an event.
\item \emph{$D(t_k)=0$, commit branch.} The branch condition gives $\gamma_c\ge N/R$, so $D'=N-R\gamma_c\le0$.
\item \emph{$D(t_k)=0$, mixing branch.} $D'=0$.
\end{itemize}
After the last event (\cref{lem:events}\ref{it:ev-b}), at most one job $j$ is unfinished and runs alone at rate $1$. All terms of $\Phi$ are non-decreasing and the diagonal term $e_j$ grows at rate $1$, so $\Phi$ grows at rate at least $1$, and $D'\le1-R<0$; after the last completion $D$ is constant. Hence $D\le0$ at all times. When the last job completes, $e=p$, so $\MRC_R=F(\infty)\le R\,\Phi(p)=R\,\OPT(p)$.
\end{proof}


\section{Consistency of $\MRC_R$}\label{app:cons}

This section proves that $\MRC_R$ is $c(R)$-consistent (\cref{thm:ft-cons}): for every $R>2$ and every instance with a correct predicted order, $\MRC_R\le c(R)\,\OPT$. Throughout, $R>2$. The proof has three parts.
\begin{itemize}
\item \Cref{app:layers} computes the cost of $\MRC_R$ on a correct order: it is $\OPT$ plus a Round-Robin overhead, $\int_0^\infty\frac{U_s(U_s-1)}2\,ds$, where $U_s$ is the number of unfinished jobs when the level reaches $s$ (\cref{cor:cost}).
\item \Cref{app:budget} bounds this overhead layer by layer, by comparing $\MRC_R$ with a \emph{target}: a function $x$ that prescribes, for each layer, how many jobs may still be unfinished when the level reaches it. For every target that satisfies the conditions \eqref{eq:target}, we get $\MRC_R\le\big(1+\int_0^1(1-x)^2\big)\OPT$ (\cref{thm:budget}).
\item \Cref{app:profile} exhibits the target $x^*$ for which this bound is $c(R)\,\OPT$ (\cref{lem:profile}), which proves \cref{thm:ft-cons}.
\end{itemize}
From the earlier sections we use the layer representation of $\Phi$ (\cref{lem:layer-phi}), the structure of the run (\cref{lem:structure,lem:phirate,lem:events}) and the invariant $D\le0$ (\cref{thm:ft-rob}).

\subsection{The cost of $\MRC_R$ on a correct order}\label{app:layers}

Fix an instance and a correct predicted order $\sigma$, and relabel the jobs so that $\sigma$ is the identity; then $p_1\le\dots\le p_n$. For $s\ge0$ let
\[
n_s=|\{i:p_i>s\}|
\]
be the number of jobs of size $>s$; it is right-continuous and non-increasing in $s$, and since the sizes are sorted, the jobs of size $>s$ are the last $n_s$ jobs, $n-n_s+1,\dots,n$.

\paragraph{Layers.} We call a value $s\ge0$ of elapsed time a \emph{layer} (think of the slice $[s,s+ds]$), and say that job $i$ \emph{crosses} layer $s$ when its elapsed time passes $s$. By \cref{lem:layer-phi}, $\Phi(e)=\int_0^\infty\Phi_s(e)\,ds$ with $\Phi_s(e)=\frac{k_s(e)(k_s(e)+1)}2$, where $k_s(e)$ is the number of jobs that have crossed layer $s$. \Cref{lem:layer-acc} below writes the cost $F$ in the same way, by charging each unit of processing to the layer at which it is given. We first record how the run looks on a correct order.

\begin{lemma}[$\MRC_R$ on a correct order]\label{lem:ft-level}
Let $R\ge2$ and consider the run of $\MRC_R$ on the relabelled instance.
\begin{enumerate}[label=(\alph*)]
\item\label{it:ftl-a} The jobs complete in the order $1,\dots,n$, at distinct times. Hence, while job $i$ is the committed job, the unfinished jobs are exactly $i,i+1,\dots,n$, and $N=n-i+1$.
\item\label{it:ftl-b} Let $t_f$ be the first time with $N(t_f)\le1$ (an event time, by \cref{lem:events}\ref{it:ev-b}), and define the \emph{level} by $\ell(t)=e_n(t)$ for $t\le t_f$ and $\ell(t)=\ell_f:=e_n(t_f)$ for $t\ge t_f$. Then $\ell$ is continuous and non-decreasing, $\ell(0)=0$, and while $N\ge2$ it is the level $\ell$ of \cref{app:run}. The level rises only in the mixing branch. At every time, the unfinished jobs other than the committed job have elapsed time $\ell(t)$, and the committed job, if any, has elapsed time at least $\ell(t)$.
\end{enumerate}
\end{lemma}
\begin{proof}
\ref{it:ftl-a}. The committed job $j$ is the first unfinished job in $\sigma$. By \cref{lem:structure}\ref{it:st-a} (which holds at all times, by \cref{lem:events}), as long as $j$ is unfinished the other unfinished jobs are at the level $\ell\le e_j<p_j$, and their sizes are at least $p_j$; so none of them completes before $j$. Nor at the same time: while $j$ is committed, $e_j-\ell$ is strictly increasing (\cref{lem:structure}\ref{it:st-a}), and $e_j-\ell\ge0$ when $j$ becomes committed, so $e_j-\ell>0$ when $j$ completes; the other unfinished jobs are then at elapsed time $\ell<e_j=p_j$, below their sizes. Hence the jobs complete in the order of $\sigma$, at distinct times, and while job $i$ is committed the completed jobs are $1,\dots,i-1$.

\ref{it:ftl-b}. While $N\ge2$, job $n$, the last in $\sigma$, is not the committed job, so by \cref{lem:structure}\ref{it:st-a} it is at the level: $\ell=e_n$, which is continuous and non-decreasing. In the commit branch the waiting jobs do not run, so the level rises only in the mixing branch. For $t\ge t_f$ the level is constant, and at most job $n$ is unfinished; if it is, it is the committed job, with elapsed time $e_n(t)\ge e_n(t_f)=\ell_f$. For $t<t_f$ the last claim is \cref{lem:structure}\ref{it:st-a}.
\end{proof}

\paragraph{The counts $U_s$.} For a value $s$ that the level reaches, let $t_s$ be the first time with $\ell(t_s)=s$, and let $U_s=N(t_s)$ be the number of unfinished jobs at that moment. If the level never reaches $s$, set $U_s=0$. The level is constant on finitely many time intervals of positive length: by \cref{lem:ft-level}\ref{it:ftl-b} these are the commit-branch intervals between consecutive events, finitely many by \cref{lem:events}, and $[t_f,\infty)$. We call $s$ \emph{regular} if it is not a job size and not one of the finitely many values taken by the level on these intervals. If a regular $s$ is reached by the level, then $\ell(u)<s$ for $u<t_s$ and $\ell(u)>s$ for $u>t_s$; moreover $s<\ell_f$, so $t_s<t_f$ and $N(t_s)\ge2$.

\begin{lemma}[Layer accounting]\label{lem:layer-acc}
Let $R\ge2$ and $t\ge0$, and write $\ell=\ell(t)$, $k_s=k_s(e(t))$ and $\Phi_s=\Phi_s(e(t))$. Then
\[
F(t)=\int_0^\infty F_s\,ds\qquad\text{and}\qquad\Phi(e(t))=\int_0^\infty\Phi_s\,ds,
\]
where, for all but finitely many $s$,
\begin{align*}
s<\ell:&\qquad \Phi_s=\tfrac{n_s(n_s+1)}2,\qquad F_s=\Phi_s+\tfrac{U_s(U_s-1)}2;\\
s>\ell:&\qquad \Phi_s=\tfrac{k_s(k_s+1)}2,\qquad F_s=k_sn_s-\tfrac{k_s(k_s-1)}2 .
\end{align*}
\end{lemma}
\begin{proof}
\emph{The formula for $\Phi$.} By \cref{lem:layer-phi}, $\Phi(e(t))=\int_0^\infty\Phi_s\,ds$ with $\Phi_s=\frac{k_s(k_s+1)}2$ for every $s$. For $s<\ell$ we have $k_s=n_s$: a job with elapsed time $>s$ has size $p_i\ge e_i>s$; conversely, a job with $p_i>s$ has elapsed time $p_i>s$ if it is complete, and at least $\ell>s$ otherwise (\cref{lem:ft-level}\ref{it:ftl-b}).

\emph{Charging the cost to layers.} For $s\in[0,e_i(t))$ let
\[
\tau_i(s)=\inf\{u:e_i(u)>s\}
\]
be the time at which job $i$ crosses layer $s$; by continuity, $e_i(\tau_i(s))=s$. By \cref{lem:events}, $[0,t]$ splits into finitely many intervals $I_1,\dots,I_M$ on each of which the rates and $N$ are constant; write $r_{i,m}$ and $N_m$ for their values on $I_m$, and $[u_m,v_m)$ for $I_m$. On $I_m$, the elapsed time $e_i$ is linear, from $e_i(u_m)$ to $e_i(v_m)=e_i(u_m)+r_{i,m}(v_m-u_m)$:
\[
e_i(s) = e_i(u_m)+r_{i,m}(s-u_m) \qquad \text{for } s \in [u_m, v_m]\;.
\]
Let $s\in[0,e_i(t)) \setminus \{e_i(u_m), e_i(v_m)\}_{m \leq M}$, then there is a unique $m$ with $e_i(u_m)<s<e_i(v_m)$. We have that $r_{i,m}>0$ and $\tau_i(s)\in I_m$, hence
\begin{equation}\label{eq:Ntau}
N\big(\tau_i(s)\big)=N_m\qquad\text{for } \,e_i(u_m)<s<e_i(v_m) .
\end{equation}
The run is non-idling (the rates sum to $a+(1-a)=1$ while $N\ge2$, and the last job runs alone at rate $1$), so $\sum_ir_{i,m}=1$ whenever $N_m\ge1$. Hence, using \eqref{eq:Ntau},
\begin{align*}
F(t)
&=\int_0^tN(u)\,du
=\sum_mN_m(v_m-u_m) \\
&=\sum_i\sum_mN_m\,r_{i,m}(v_m-u_m)\\
&=\sum_i\sum_m\int_{e_i(u_m)}^{e_i(v_m)}N_m\,ds\\
&=\sum_i\int_0^{e_i(t)}N\big(\tau_i(s)\big)\,ds\\
&=\int_0^\infty F_s\,ds,
\end{align*}
where
\[
F_s=\sum_{i:\,e_i(t)>s}N\big(\tau_i(s)\big)
\]
is the sum, over the jobs that have crossed layer $s$, of the number of unfinished jobs at the moment they crossed it. It remains to compute $F_s$ for regular $s\ne\ell$ outside the finite set of values $e_i(u_m),e_i(v_m)$.

\emph{Crossing while committed.} If job $i$ crosses layer $s$ at time $\tau=\tau_i(s)$ while it is the committed job, then $N(\tau)=n-i+1$ by \cref{lem:ft-level}\ref{it:ftl-a}. Writing the jobs of size $>s$ as $n-n_s+r$, $r=1,\dots,n_s$, the $r$-th of them contributes $N(\tau)=n_s-r+1$. A job $i$ that crosses $s$ at a time $\tau$ with $\ell(\tau)<s$ does so while committed: it is unfinished at $\tau$ (as $e_i(\tau)=s<e_i(t)\le p_i$) and above the level, whereas the other unfinished jobs are at the level (\cref{lem:ft-level}\ref{it:ftl-b}).

\emph{Layers $s>\ell$.} Let $i$ be one of the $k_s$ jobs with $e_i(t)>s$; then $p_i>s$, and $i$ crossed $s$ at $\tau_i(s)<t$, when the level was $\ell(\tau_i(s))\le\ell<s$; so it crossed while committed. If $i'<i$ is another job of size $>s$, then $i'$ was complete when $i$ was committed (\cref{lem:ft-level}\ref{it:ftl-a}), so $e_{i'}(t)=p_{i'}>s$. Hence the $k_s$ jobs that crossed layer $s$ are the first $k_s$ jobs of size $>s$, and they contribute $n_s,n_s-1,\dots,n_s-k_s+1$:
\[
F_s=\sum_{r=1}^{k_s}(n_s-r+1)=k_sn_s-\frac{k_s(k_s-1)}2 .
\]

\emph{Layers $s<\ell$.} The level is continuous with $\ell(0)=0\le s<\ell(t)$, so it reached $s$, at time $t_s<t$. As shown above, all $n_s$ jobs of size $>s$ have crossed layer $s$. The $U_s=N(t_s)$ jobs unfinished at $t_s$ are the last ones, $n-U_s+1,\dots,n$ (\cref{lem:ft-level}\ref{it:ftl-a}).
\begin{itemize}
\item The jobs $i=n-n_s+1,\dots,n-U_s$ are complete at $t_s$. Each crossed $s$ at a time $\tau<C_i\le t_s$ (as $e_i(C_i)=p_i>s$ and $e_i$ is continuous), when $\ell(\tau)<s$, hence while committed, and contributes $n-i+1$. Together they contribute $n_s+(n_s-1)+\dots+(U_s+1)=\frac{n_s(n_s+1)}2-\frac{U_s(U_s+1)}2$.
\item The waiting jobs at time $t_s$ have elapsed time $s$. At every time $u>t_s$, each of them has elapsed time $>s$: at least $\ell(u)>s$ if it is unfinished, and its size if it is complete, which exceeds $s$ since the job was unfinished at $t_s$ with elapsed time $s$. So they cross layer $s$ at time $t_s$, and each contributes $N(t_s)=U_s$.
\item The committed job $n-U_s+1$ at time $t_s$ has elapsed time $\ge s$. If it equals $s$, the job crosses at time $t_s$, for the same reason as the waiting jobs; otherwise it crossed at a time $\tau<t_s$, while committed, and contributes $n-(n-U_s+1)+1=U_s$. In both cases it contributes $U_s$.
\end{itemize}
Adding up,
\begin{align*}
F_s
&=\frac{n_s(n_s+1)}2-\frac{U_s(U_s+1)}2+U_s\cdot U_s \\
&=\frac{n_s(n_s+1)}2+\frac{U_s(U_s-1)}2 .
\qedhere
\end{align*}
\end{proof}

\begin{corollary}[Cost of $\MRC_R$]\label{cor:cost}
Let $R\ge2$. On a correct order,
\begin{equation}\label{eq:cost}
\MRC_R=\OPT+\int_0^\infty\frac{U_s(U_s-1)}2\,ds .
\end{equation}
\end{corollary}
The $U_s$ jobs still unfinished when the level reaches $s$ cross layer $s$ together, in Round-Robin. In SPT, of each pair of these jobs only one waits for the other while the layer is processed; in Round-Robin both wait. So each of the $\frac{U_s(U_s-1)}2$ pairs pays $ds$ extra at layer $s$, and this is the whole excess cost of $\MRC_R$ over $\OPT$.
\begin{proof}
Let $t$ be a time at which all jobs are complete. Then $\MRC_R=F(t)$, $e(t)=p$, and $\ell(t)=\ell_f$. Apply \cref{lem:layer-acc} at time $t$. For $s<\ell_f$, $F_s=\frac{n_s(n_s+1)}2+\frac{U_s(U_s-1)}2$. For $s>\ell_f$, every job has elapsed time $p_i$, so $k_s=n_s$ and $F_s=n_s^2-\frac{n_s(n_s-1)}2=\frac{n_s(n_s+1)}2$; moreover the level never reaches $s$, so $U_s=0$. So $F_s=\frac{n_s(n_s+1)}2+\frac{U_s(U_s-1)}2$ for all but finitely many $s$. Integrating, and using $\int_0^\infty\frac{n_s(n_s+1)}2ds=\Phi(p)=\OPT$ (\cref{lem:layer-phi} with $e=p$, for which $k_s(p)=n_s$), gives 
\[
\MRC_R=\OPT+\int_0^\infty\frac{U_s(U_s-1)}2\,ds \;.
\]
\end{proof}

\subsection{Comparison with a target}\label{app:budget}

For $\xi\in[0,1]$ let
\begin{equation}\label{eq:hq}
h(\xi)=\xi-\frac{1+R}2\xi^2,\qquad q(\xi)=\frac{(1-\xi)^2-(R-1)}2,\qquad \tilde{\xi}=\frac1{1+R}.
\end{equation}
The function $h$ is concave, with $h'(\xi)=1-(1+R)\xi$; it is increasing on $[0,\tilde{\xi}]$, decreasing on $[\tilde{\xi},1]$, and $h(\tilde{\xi})=\frac1{2(1+R)}$ is its maximum. Also $q(1)=-\frac{R-1}2$.

\begin{theorem}[Comparison with a target]\label{thm:budget}
Let $R>2$. Let $x\colon[0,1]\to[0,1]$ be non-decreasing and right-continuous, and let $\tilde{w}\in(0,1]$ be such that $x(w)<\tilde{\xi}$ for $w<\tilde{w}$ and $x(w)=1$ for $w\ge\tilde{w}$. For $w\in[0,\tilde{w})$ let
\[
\Gamma_x(w)=\int_0^wq\big(x(v)\big)\,dv+(1-w)\,h\big(x(w)\big),
\]
and suppose that
\begin{equation}\label{eq:target}
\Gamma_x(w)\le0\ \text{ for all }w\in[0,\tilde{w}),\qquad\text{and}\qquad \int_0^{\tilde{w}}q\big(x(v)\big)\,dv+(1-\tilde{w})\,h(\tilde{\xi})\le0 .
\end{equation}
Then, on every instance with a correct predicted order,
\[
\MRC_R\ \le\ \Big(1+\int_0^1\big(1-x(w)\big)^2dw\Big)\OPT .
\]
\end{theorem}
The function $x$ is a \emph{target}: it prescribes, in a coordinate $w\in[0,1]$ defined in Step~1 below, the fraction of jobs that should have been completed when the level is at $w$. On $n$ jobs of size $1$, a schedule that commits in this way has cost $\big(1+\int_0^1(1-x)^2\big)\OPT$ and $D / \OPT=2\,\Gamma_x(w)$ when the level is at $w$, up to $O(1/n)$ (\cref{sec:overview}); so \eqref{eq:target} says that this schedule keeps $D\le0$. $\MRC_R$ does not know $x$ or the sizes. The proof shows that, at the first time it would fall behind the target, committing would keep $D\le0$, so it would have committed.
\begin{proof}
Relabel the jobs as in \cref{app:layers}, so that $p_1\le\dots\le p_n$, and let $\ell(t)$ be the level and $U_s$ the counts of the run of $\MRC_R$. The \emph{branch at time $t$} is the branch chosen at the last event time $t_k\le t$; it is used on all of $[t_k,t_{k+1})$. By \cref{lem:layer-phi} with $e=p$, we can write
\[
\OPT=\Phi(p)=\int_0^\infty\Phi_s(p)\,ds,\qquad \text{ with } \qquad \Phi_s(p)=\frac{n_s(n_s+1)}2 \;.
\]
\noindent\textbf{Step 1: the coordinate $w$ and the budget.} For $s\ge0$ let
\[
w(s)=\frac1{\OPT}\int_0^s\Phi_v(p)\,dv
\]
be the fraction of $\OPT$ contributed by the layers below $s$. It is continuous and non-decreasing, with $w(0)=0$ and $w(s)=1$ for $s\ge p_n$, and $w'(s)=\Phi_s(p)/\OPT$ for every $s$ that is not a job size. So $\Phi_s(p)\,ds=\OPT\,dw$: for every bounded Borel function $g$ on $[0,1]$ and every $\ell\ge0$, the change of variables $v=w(s)$ gives
\begin{equation}\label{eq:cov}
\int_0^\ell\Phi_s(p)\,g\big(w(s)\big)\,ds=\OPT\int_0^{w(\ell)}g(v)\,dv,\qquad\text{and}\qquad \int_\ell^\infty\Phi_s(p)\,ds=\big(1-w(\ell)\big)\OPT .
\end{equation}
For $s\ge0$ we define the \emph{budget} as
\[
T_s=\Phi_s(p)\,\big(1-x(w(s))\big)^2 .
\]
We will prove that
\begin{equation}\label{eq:dom}
\frac{U_s(U_s-1)}2\le T_s\qquad\text{for all but finitely many }s ,
\end{equation}
which gives the theorem by \cref{cor:cost} and \eqref{eq:cov}. The function $s\mapsto T_s$ is non-increasing, since $\Phi_s(p)$ is non-increasing (as $n_s$ is) and $x\circ w$ is non-decreasing with values in $[0,1]$. It is right-continuous, since $\Phi_s(p)$ is (as $n_s$ is) and $x\circ w$ is: if $s\downarrow s_0$ then $w(s)\downarrow w(s_0)$, and $x$ is right-continuous.

\noindent\textbf{Step 2: the first violation.} Let $\mathcal T$ be the set of times $t$ at which $N(t)\ge2$, the branch is mixing, and
$\frac{N(t)(N(t)-1)}2\ge T_{\ell(t)}$.
We show two things: if $\mathcal T=\emptyset$ then \eqref{eq:dom} holds, and if $\mathcal T\ne\emptyset$ then it has a least element $t^*$. Steps~3--7 then show that such a $t^*$ cannot exist.
\begin{itemize}
    \item \emph{If $\mathcal T=\emptyset$, then \eqref{eq:dom} holds.} Let $s$ be regular. If the level never reaches $s$, then $U_s=0\le T_s$. Otherwise it reaches $s$ at time $t_s$, with $t_s<t_f$ and $N(t_s)=U_s\ge2$. Let $t_k\le t_s<t_{k+1}$ be consecutive event times. The level exceeds $s$ just after $t_s$, so it is not constant on $[t_k,t_{k+1})$; since it does not move in the commit branch (\cref{lem:ft-level}\ref{it:ftl-b}), the branch at $t_s$ is mixing. So $t_s\notin\mathcal T$ means exactly $\frac{U_s(U_s-1)}2<T_s$.
    \item \emph{If $\mathcal T\ne\emptyset$, it has a least element.} Let $t^*=\inf\mathcal T$, and let $t_k\le t^*<t_{k+1}$ be consecutive event times. Since $t^*$ is the infimum, $\mathcal T$ meets $[t^*,t^*+\eta)$ for every $\eta>0$, hence meets $[t_k,t_{k+1})$, on which $N$ and the branch are constant; so $N(t^*)\ge2$ and the branch at $t^*$ is mixing. The function $t\mapsto T_{\ell(t)}$ is non-increasing and right-continuous on $[t_k,t_{k+1})$, since $\ell$ is continuous and non-decreasing and $T$ is non-increasing and right-continuous. Taking $t\downarrow t^*$ along times of $\mathcal T$ in $[t_k,t_{k+1})$, where $N(t)=N(t^*)$,
\[
T_{\ell(t^*)}=\lim_{t\downarrow t^*}T_{\ell(t)}\ \le\ \frac{N(t^*)(N(t^*)-1)}2 .
\]
Therefore, it holds that $t^*\in\mathcal T$.
\end{itemize}

Suppose, for a contradiction, that $\mathcal T\ne\emptyset$. The rest of the proof shows that committing at time $t^*$ would keep $D\le0$, whereas $\MRC_R$ mixes at $t^*$, which means that committing alone would make $D$ positive.

\paragraph{Step 3: the state at $t^*$.} Let $\ell=\ell(t^*)$ and $u=N(t^*)\ge2$. By \cref{lem:ft-level}, the committed job is $j=n-u+1$, the unfinished jobs are $j,\dots,n$, the waiting jobs have elapsed time $\ell$, and $z:=e_j(t^*)\in[\ell,p_j)$. We use three facts about this time.
\begin{itemize}
\item \emph{The past is within budget:} $\frac{U_s(U_s-1)}2<T_s$ for every regular $s<\ell$. Indeed, the level reached $s$ at a time $t_s<t^*$, since $\ell$ is continuous and non-decreasing; as in Step~2, the branch at $t_s$ is mixing and $N(t_s)\ge2$, and $t_s\notin\mathcal T$ by minimality of $t^*$.
\item \emph{The present is over budget:} $\frac{u(u-1)}2\ge T_\ell$, since $t^*\in\mathcal T$.
\item \emph{Committing alone is too slow:} $D(t^*)=0$ and $\gamma_c<u/R$, with $\gamma_c$ evaluated at $t^*$. Indeed, let $t_k\le t^*<t_{k+1}$ be consecutive event times. The mixing branch was chosen at $t_k$, so the commit condition failed there: $D(t_k)\ge0$ and $\gamma_c(t_k)<N(t_k)/R$. By \cref{thm:ft-rob}, $D(t_k)\le0$, so $D(t_k)=0$; by \cref{lem:events}\ref{it:ev-a}, $D'=0$ on $[t_k,t_{k+1})$, so $D(t^*)=0$, and $\gamma_c(t^*)=\gamma_c(t_k)<u/R$.
\end{itemize}

\noindent\textbf{Step 4: a hypothetical commit.} Suppose that, from time $t^*$, the algorithm ran $j$ alone until its elapsed time reached $z'$, for some $z'\in[z,p_j]$. The elapsed times would become $e'$, equal to $e(t^*)$ except $e'_j=z'$. All $u$ jobs would stay unfinished during that time ($j$ completes at the end if $z'=p_j$), so the cost so far would become $F(t^*)+u(z'-z)$. Let
\[
\psi(z')=F(t^*)+u(z'-z)-R\,\Phi(e')
\]
be the value that $D$ would have after this commit. Then $\psi(z)=D(t^*)=0$. Raising $e_j$ from $z$ at rate $1$ is committing from the state at time $t^*$, so by \cref{lem:phirate} (with $a=1$) $z'\mapsto\Phi(e')$ has right derivative $\gamma_c$ at $z'=z$, and $\psi$ has right derivative $u-R\gamma_c>0$ there (Step~3). Since $z<p_j$, $\psi(z')>0$ for some $z'\in(z,p_j]$. We now show that $\psi(z')\le0$ for every $z'\in[z,p_j]$, which is a contradiction.

We write $\psi(z')$ layer by layer. Let $k'_s=k_s(e')$ be the number of jobs with elapsed time $>s$ after the commit, and $k_s=k_s(e(t^*))$ before it. By \cref{lem:layer-phi}, $\Phi(e')=\int_0^\infty\Phi_s(e')\,ds$, and by \cref{lem:layer-acc} at time $t^*$, $F(t^*)=\int_0^\infty F_s\,ds$. The commit changes only the layers $s\in[z,z')$: there $k'_s=k_s+1$, and $u=n_s-k_s$, because the jobs of size $>s$ are the $u$ unfinished jobs (all of size $\ge p_j>s$) and the $n_s-u$ complete jobs of size $>s$, while $j$ and the waiting jobs have elapsed time $\le s$. So $u(z'-z)=\int_z^{z'}(n_s-k_s)\,ds$, and on $[z,z')$
\begin{align*}
k_sn_s-\frac{k_s(k_s-1)}2+(n_s-k_s)
&=(k_s+1)n_s-\frac{(k_s+1)k_s}2\\
&=k'_sn_s-\frac{k'_s(k'_s-1)}2 .
\end{align*}

For $s<\ell$ the commit changes nothing, and $k'_s=k_s=n_s$. Hence, with the formulas of \cref{lem:layer-acc},
\begin{equation}\label{eq:psi-layers}
\psi(z')=\int_0^\ell\Big(\frac{U_s(U_s-1)}2-(R-1)\Phi_s(p)\Big)ds+\int_\ell^\infty\Big(k'_sn_s-\frac{k'_s(k'_s-1)}2-R\,\frac{k'_s(k'_s+1)}2\Big)ds .
\end{equation}
The first integral depends only on the past, the second only on the state after the commit. We bound them in Steps~5 and~6.

\noindent\textbf{Step 5: the layers below the level.} For every regular $s<\ell$, the past is within budget (Step~3), so the integrand of the first integral in \eqref{eq:psi-layers} is at most
\begin{align*}
T_s-(R-1)\Phi_s(p)
&=\Phi_s(p)\Big[\big(1-x(w(s))\big)^2-(R-1)\Big]\\
&=2\,\Phi_s(p)\,q\big(x(w(s))\big).    
\end{align*}

The function $q\circ x$ is bounded and Borel ($x$ is monotone), so \eqref{eq:cov} with $g=q\circ x$ bounds the first integral by
\[
2\,\OPT\int_0^{w(\ell)}q\big(x(v)\big)\,dv .
\]

\noindent\textbf{Step 6: the layers above the level.} Let $\xi_\ell=x(w(\ell))$ and
\[
H=\begin{cases}h(\xi_\ell)&\text{if }w(\ell)<\tilde{w}\quad\text{(then $\xi_\ell<\tilde{\xi}$)},\\ h(\tilde{\xi})&\text{if }w(\ell)\ge \tilde{w}.\end{cases}
\]
In both cases $H\ge h(0)=0$, as $h$ is increasing on $[0,\tilde{\xi}]$. We show that, for every $s\ge\ell$, the integrand of the second integral in \eqref{eq:psi-layers} is at most $n_s(n_s+1)H=2\Phi_s(p)H$.

Fix $s\ge\ell$ and write $m=n_s$ and $k=k'_s$. After the commit, a complete job $i$ has elapsed time $p_i$, the waiting jobs have elapsed time $\ell\le s$, and $j$ has elapsed time $z'\le p_j$.
\begin{itemize}
\item If $u>m$, then $j$, the smallest unfinished job, is not among the last $m$ jobs, so $p_j\le s$; every complete job $i<j$ has $p_i\le p_j\le s$, and $z'\le p_j\le s$. So $k=0$, the integrand is $0$, and $0\le2\Phi_s(p)H$.
\item If $u\le m$, the $u$ unfinished jobs are among the $m$ jobs of size $>s$, so $m-u$ of these are complete; the waiting jobs have elapsed time $\ell\le s$, so $k\le m-u+1$. If $k=0$ we conclude as before. If $k\ge1$, let $\xi'=k/m$, then $\xi'\le\frac{m-u+1}m=1-\frac{u-1}m$. Since $k(k-1)+Rk(k+1)=(1+R)k^2+(R-1)k$, it follows that
\[
km-\frac{k(k-1)}2-R\,\frac{k(k+1)}2=m^2h(\xi')-\frac{(R-1)k}2\le m^2h(\xi') .
\]
So it remains to show $m^2h(\xi')\le m(m+1)H$ in this case.
\end{itemize}
If $w(\ell)\ge\tilde{w}$, then $H=h(\tilde{\xi})=\max h>0$, so
\[
m^2h(\xi')\le m^2h(\tilde{\xi})\le m(m+1)\,h(\tilde{\xi})=2\Phi_s(p)H\;.
\]
Otherwise, if $w(\ell)<\tilde{w}$, then $H=h(\xi_\ell)$ with $\xi_\ell<\tilde{\xi}$, and we show $m^2h(\xi')\le m(m+1)\,h(\xi_\ell)$, using that the present is over budget: $\frac{u(u-1)}2\ge T_\ell=(1-\xi_\ell)^2\frac{n_\ell(n_\ell+1)}2$. We have that
\[
\Big(u-\frac12\Big)^2=u(u-1)+\frac14\ \ge\ (1-\xi_\ell)^2n_\ell(n_\ell+1)+\frac{(1-\xi_\ell)^2}4\ =\ (1-\xi_\ell)^2\Big(n_\ell+\frac12\Big)^2 .
\]
Both $u-\frac12$ and $(1-\xi_\ell)(n_\ell+\frac12)$ are non-negative, so $u-\frac12\ge(1-\xi_\ell)(n_\ell+\frac12)$, i.e.
\[
u-1\ \ge\ (1-\xi_\ell)\,n_\ell-\frac{\xi_\ell}2 .
\]
Since $m=n_s\le n_\ell$ (as $s\ge\ell$), $(1-\xi_\ell)n_\ell\ge(1-\xi_\ell)m$, and dividing by $m$,
\[
\xi'\ \le\ 1-\frac{u-1}m\ \le\ \xi_\ell+\frac{\xi_\ell}{2m} .
\]
If $\xi'\le\xi_\ell$, then $h(\xi')\le h(\xi_\ell)$, as $h$ is increasing on $[0,\tilde{\xi}]$. Otherwise $0<\xi'-\xi_\ell\le\frac{\xi_\ell}{2m}$; since $h$ is concave it lies below its tangent at $\xi_\ell$, and $h'(\xi_\ell)=1-(1+R)\xi_\ell\ge0$, so
\begin{align*}
h(\xi')&\le h(\xi_\ell)+h'(\xi_\ell)\,\frac{\xi_\ell}{2m}=h(\xi_\ell)+\frac{\xi_\ell-(1+R)\xi_\ell^2}{2m}\\
&\le h(\xi_\ell)+\frac{2\xi_\ell-(1+R)\xi_\ell^2}{2m}\\
&=\Big(1+\frac1m\Big)h(\xi_\ell),
\end{align*}
using $2\xi_\ell-(1+R)\xi_\ell^2=2h(\xi_\ell)$. In both cases $m^2h(\xi')\le m^2(1+\frac1m)h(\xi_\ell)=m(m+1)h(\xi_\ell)$.

So in both cases the integrand is at most $2\Phi_s(p)H$ for every $s\ge\ell$, and by \eqref{eq:cov} the second integral in \eqref{eq:psi-layers} is at most $2\,\OPT\,(1-w(\ell))\,H$.

\noindent\textbf{Step 7: committing would be safe.} Adding the bounds of Steps~5 and~6,
\[
\frac{\psi(z')}{2\,\OPT}\le\int_0^{w(\ell)}q\big(x(v)\big)\,dv+\big(1-w(\ell)\big)H .
\]
If $w(\ell)<\tilde{w}$, then $H=h(x(w(\ell)))$ and the right side is $\Gamma_x(w(\ell))\le0$ by \eqref{eq:target}. If $w(\ell)\ge\tilde{w}$, then $H=h(\tilde{\xi})$, and $q(x(v))=q(1)=-\frac{R-1}2$ for $v\ge\tilde{w}$; splitting the integral at $\tilde{w}$ and writing $1-w(\ell)=(1-\tilde{w})-(w(\ell)-\tilde{w})$, the right side equals
\begin{align*}
&\int_0^{\tilde{w}}q\big(x(v)\big)\,dv+(1-\tilde{w})\,h(\tilde{\xi})-\big(w(\ell)-\tilde{w}\big)\Big(\frac{R-1}2+h(\tilde{\xi})\Big)\\
&\qquad\le-\big(w(\ell)-\tilde{w}\big)\Big(\frac{R-1}2+h(\tilde{\xi})\Big)\le0
\end{align*}
by \eqref{eq:target}. In both cases $\psi(z')\le0$ for every $z'\in[z,p_j]$, which contradicts Step~4. Hence $\mathcal T=\emptyset$, and \eqref{eq:dom} holds by Step~2.

\noindent\textbf{Step 8: the cost.} By \cref{cor:cost}, \eqref{eq:dom}, and \eqref{eq:cov} with $\ell=p_n$, $w(p_n)=1$ and $g(v)=(1-x(v))^2$, and noting that $T_s=0$ for $s\ge p_n$ (where $n_s=0$), we deduce that
\begin{align*}
\MRC_R&=\OPT+\int_0^\infty\frac{U_s(U_s-1)}2\,ds\\
&\le\OPT+\int_0^{p_n}\Phi_s(p)\big(1-x(w(s))\big)^2ds\\
&=\OPT\Big(1+\int_0^1\big(1-x(v)\big)^2dv\Big).
\qedhere
\end{align*}
\end{proof}

\subsection{The profile}\label{app:profile}

In this section we define a target $x^*$, show that it satisfies the conditions of \cref{thm:budget}, and that $\int_0^1(1-x^*)^2=c(R)-1$, which proves the consistency bound. Recall that $\tilde{\xi} = 1/(R+1)$, and let
\begin{equation}\label{eq:consts}
\beta=\frac{R-2}{R+2},\qquad \gamma=\frac R{R+2},\qquad \alpha=1-\gamma=\frac2{R+2},\qquad K=1+\frac{\tilde{\xi}}\beta=\frac{R^2}{(R+1)(R-2)} \;.
\end{equation}
All of $\beta,\gamma,\alpha,\tilde{\xi}$ lie in $(0,1)$, and $K>1$. Define on $[0,\tilde{\xi}]$
\begin{equation}\label{eq:W}
W(\xi)=1-\frac1{(1-\xi)\,(1+\xi/\beta)^{\gamma}},\qquad \tilde{w}=W(\tilde{\xi})=1-\frac{1}{(1-\tilde{\xi})K^{\gamma}} ,
\end{equation}
the \emph{profile}
\[
x^*(w)=W^{-1}(w)\quad\text{for }w\in[0,\tilde{w}),\qquad x^*(w)=1\quad\text{for }w\in[\tilde{w},1],
\]
and
\begin{equation}\label{eq:lambda}
\lambda(R)=R-1-\frac{(R+1)(R-2)}R\,K^{\alpha} .
\end{equation}

\begin{lemma}[The profile]\label{lem:profile} With the definitions above, it holds that
\begin{enumerate}[label=(\alph*)]
\item\label{it:pa} $W$ is $C^1$ and strictly increasing on $[0,\tilde{\xi}]$, with $W(0)=0$, $0<\tilde{w}<1$ and
\[
W'(\xi)=\frac{2\,(1-W(\xi))\,(1-(1+R)\xi)}{(R+2)(1-\xi)(\xi+\beta)} .
\]
Consequently $x^*$ is well defined, non-decreasing and right-continuous, with $x^*<\tilde{\xi}$ on $[0,\tilde{w})$.
\item\label{it:pb} $x^*$ satisfies \eqref{eq:target} with equality: $\Gamma_{x^*}=0$ on $[0,\tilde{w})$, and $\int_0^{\tilde{w}}q(x^*(v))\,dv+(1-\tilde{w})\,h(\tilde{\xi})=0$.
\item\label{it:pc} $\int_0^1(1-x^*(w))^2\,dw=\lambda(R)=c(R)-1$.
\end{enumerate}
\end{lemma}
\begin{proof}
\ref{it:pa}. On $[0,\tilde{\xi}]$ we have $1-\xi>0$ and $1+\xi/\beta>0$, so $1-W(\xi)=(1-\xi)^{-1}(1+\xi/\beta)^{-\gamma}>0$ and $W$ is $C^\infty$. Taking logarithms
\[
\ln(1-W(\xi))=-\ln(1-\xi)-\gamma\ln(1+\xi/\beta)\;,
\]
then differentiating gives
\[
\frac{W'(\xi)}{1-W(\xi)}=\frac\gamma{\xi+\beta}-\frac1{1-\xi}=\frac{\gamma-\beta-(1+\gamma)\xi}{(1-\xi)(\xi+\beta)} .
\]
Since $\gamma-\beta=\frac2{R+2}$ and $1+\gamma=\frac{2(R+1)}{R+2}$, the numerator is $\frac{2(1-(1+R)\xi)}{R+2}$, which gives the formula for $W'$. On $[0,\tilde{\xi})$ every factor of $W'$ is positive, and $W'(\tilde{\xi})=0$; so $W$ is strictly increasing on $[0,\tilde{\xi}]$, and $0=W(0)<\tilde{w}$. Also $\tilde{w}<1$ because $1-\tilde{w}=\frac1{(1-\tilde{\xi})K^\gamma}>0$. Hence $W$ is a continuous increasing bijection $[0,\tilde{\xi}]\to[0,\tilde{w}]$, so $x^*=W^{-1}$ is well defined, continuous and increasing on $[0,\tilde{w})$, with values in $[0,\tilde{\xi})$. At $\tilde{w}$ it jumps to $1$ and stays there. So $x^*$ is non-decreasing and right-continuous.

\ref{it:pb}. For $\xi\in[0,\tilde{\xi}]$ let
\[
J(\xi)=\int_0^\xi q(\eta)\,W'(\eta)\,d\eta+\big(1-W(\xi)\big)\,h(\xi) \;.
\]
We have $J(0)=h(0)=0$, and $J$ is $C^1$ with
\[
J'(\xi)=W'(\xi)\big(q(\xi)-h(\xi)\big)+\big(1-W(\xi)\big)h'(\xi).
\]
Expanding, and observing that $(\xi-1)(\xi+\beta)=\xi^2-(1-\beta)\xi-\beta$ with $1-\beta=\frac4{R+2}$ and $\beta=\frac{R-2}{R+2}$, we deduce
\begin{align*}
q(\xi)-h(\xi)&=\frac{(1-\xi)^2-(R-1)}2-\xi+\frac{1+R}2\xi^2\\
&=\frac{(R+2)\xi^2-4\xi-(R-2)}2\\
&=\frac{R+2}2(\xi-1)(\xi+\beta) \;.
\end{align*}
With the formula for $W'$ from \ref{it:pa},
\begin{align*}
W'(\xi)\big(q(\xi)-h(\xi)\big)
&=\frac{2\,(1-W(\xi))\,(1-(1+R)\xi)}{(R+2)(1-\xi)(\xi+\beta)}\cdot\frac{R+2}2(\xi-1)(\xi+\beta)\\
&=-\big(1-W(\xi)\big)\big(1-(1+R)\xi\big) \\
&= -(1-W(\xi))h'(\xi)\;,
\end{align*}
therefore $J' = 0$ and $J = 0$ on $[0,\tilde{\xi}]$.

Now let $w\in[0,\tilde{w}]$ and $\xi=W^{-1}(w)\in[0,\tilde{\xi}]$. Since $W$ is a $C^1$ increasing bijection $[0,\tilde{\xi}]\to[0,\tilde{w}]$ and $x^*(W(\eta))=\eta$ for $\eta\in[0,\tilde{\xi})$, the change of variables $v=W(\eta)$ gives
\[
\int_0^wq\big(x^*(v)\big)\,dv=\int_0^{\xi}q(\eta)\,W'(\eta)\,d\eta,\qquad\text{and}\qquad 1-w=1-W(\xi).
\]
For $w<\tilde{w}$, $\xi=x^*(w)$ and $\Gamma_{x^*}(w)=J(\xi)=0$. For $w=\tilde{w}$, $\xi=\tilde{\xi}$ and
\[
\int_0^{\tilde{w}}q\big(x^*(v)\big)\,dv+(1-\tilde{w})\,h(\tilde{\xi})=J(\tilde{\xi})=0 .
\]

\ref{it:pc}. Since $x^*=1$ on $[\tilde{w},1]$, only $[0,\tilde{w})$ contributes. Substituting $w=W(\xi)$, $dw=W'(\xi)\,d\xi$, and integrating by parts with $W'=\frac{d}{d\xi}\big(-(1-W)\big)$, $W(0)=0$ and $W(\tilde{\xi})=\tilde{w}$,
\begin{align*}
\int_0^1(1-x^*(w))^2\,dw&=\int_0^{\tilde{\xi}}(1-\xi)^2\,W'(\xi)\,d\xi\\
&=\Big[-(1-\xi)^2(1-W(\xi))\Big]_0^{\tilde{\xi}}-2\int_0^{\tilde{\xi}}(1-\xi)(1-W(\xi))\,d\xi\\
&=1-(1-\tilde{\xi})^2(1-\tilde{w})-2\int_0^{\tilde{\xi}}(1-\xi)(1-W(\xi))\,d\xi .
\end{align*}
By definition of $W$, $(1-\xi)(1-W(\xi))=(1+\xi/\beta)^{-\gamma}$. At $\xi=\tilde{\xi}$, where $1+\tilde{\xi}/\beta=K$, this gives $(1-\tilde{\xi})^2(1-\tilde{w})=(1-\tilde{\xi})K^{-\gamma}$. For the integral, since $1-\gamma=\alpha$,
\[
\int_0^{\tilde{\xi}}(1+\xi/\beta)^{-\gamma}d\xi=\Big[\frac\beta{\alpha}(1+\xi/\beta)^{\alpha}\Big]_0^{\tilde{\xi}}=\frac\beta{\alpha}(K^\alpha-1),\qquad \frac{2\beta}\alpha=R-2 .
\]
Therefore, using
\[
K^{-\gamma}=\frac{K^{\alpha}}K\qquad\text{and}\qquad\frac{1-\tilde{\xi}}{K}=\frac{R}{R+1}\cdot\frac{(R+1)(R-2)}{R^2}=\frac{R-2}R ,
\]
we obtain that
\begin{align*}
\int_0^1(1-x^*(w))^2\,dw&=1-(1-\tilde{\xi})K^{-\gamma}-(R-2)(K^{\alpha}-1)\\
&=R-1-K^\alpha\Big(\frac{R-2}R+R-2\Big)\\
&=R-1-\frac{(R+1)(R-2)}R\,K^\alpha\\
&=\lambda(R) \;.
\end{align*}
Finally, $K^{-1}=\frac{(R+1)(R-2)}{R^2}=1-\frac{R+2}{R^2}$, hence
\[
\frac{(R+1)(R-2)}R\,K^{\alpha}=R\,K^{\alpha-1}=R\,K^{-\gamma}=R\Big(1-\frac{R+2}{R^2}\Big)^{\frac R{R+2}},
\]
and $\lambda(R)=c(R)-1$ by \eqref{eq:cR}.
\end{proof}

\begin{theorem}[Consistency of $\MRC_R$]\label{thm:ft-cons}
Let $R>2$ and let $c(R)$ be given by \eqref{eq:cR}. If $\sigma$ is correct, then $\MRC_R\le c(R)\,\OPT$. 
\end{theorem}
\begin{proof}
By \cref{lem:profile}\ref{it:pa} and \ref{it:pb}, $x^*$ satisfies the hypotheses of \cref{thm:budget}, with the $\tilde{w}$ of \eqref{eq:W}. So $\MRC_R\le\big(1+\int_0^1(1-x^*)^2\big)\OPT=c(R)\,\OPT$ by \cref{lem:profile}\ref{it:pc}.
\end{proof}

\section{The lower bound}\label{app:lbsec}

This section proves the lower bound (\cref{thm:lb}): for every $R>2$, any deterministic $R$-robust algorithm has consistency at least $c(R)$, given by \eqref{eq:cR}, even if it receives the predicted sizes and not only their order. 

\paragraph{Plan.} \Cref{app:freeze} proves that the robustness invariant is necessary (\cref{lem:freeze-lb}): an $R$-robust deterministic algorithm must keep $F\le R\Phi$ along every run, in particular along its run on correct predictions. \Cref{app:lb} applies it to the run on $n$ jobs of size $1$, turns the resulting constraints into a recursion, and lets $n\to\infty$.
In the proof, we reuse the constants $\tilde{\xi},\alpha,\beta,\gamma,K$ of \eqref{eq:hq} and \eqref{eq:consts}, and the identity $\lambda(R)=c(R)-1$ of \cref{lem:profile}.

\paragraph{Predicted sizes.} In this model the prediction is a vector $y\in(0,\infty)^n$, correct if $y=p$. The algorithm knows $n$ and $y$, and its consistency and robustness are defined as in \cref{app:prelim}, with $y$ in place of $\sigma$. Given an algorithm $A_\sigma$ for predicted orders, sorting $y$ (ties by index) and running $A_\sigma$ gives an algorithm $A_y$ for predicted sizes, and since a correct $y$ induces a correct order,
\[
\cons(A_y)\le\cons(A_\sigma)\qquad\text{and}\qquad\rob(A_y)\le\rob(A_\sigma) .
\]
So it suffices to prove the lower bound for predicted sizes. Below, the prediction $y$ may be an order or a vector of sizes.

\subsection{The robustness invariant is necessary}\label{app:freeze}

\begin{lemma}[Necessity of the robustness invariant]\label{lem:freeze-lb}
Let $A$ be a deterministic algorithm, $(p,y)$ any instance (sizes and predictions), and $t\ge0$. Let $e(t)$ and $F(t)$ be the elapsed times and the cost so far of the run of $A$ on $(p,y)$. If $\Phi(e(t))>0$, then
\[
\rob(A)\;\ge\;\frac{F(t)}{\Phi(e(t))}.
\]
\end{lemma}
The proof generalizes the adversary of Wei and Zhang~\citep[Theorem~1.3]{WZ20}, who end the unfinished jobs at the first completion. In \cref{app:lb} we apply it to the \emph{perfect run}, the run on $(y,y)$ in which the predictions are correct.
\begin{proof}
For a time $t$, the idea is to consider the adversarial instance where every unfinished job ends right after time $t$.
Fix $\delta>0$ and define the sizes $p^\delta$ by
\[
p^\delta_j=\begin{cases}p_j&\text{if $j$ is complete at time $t$, i.e.\ $C_j\le t$},\\ e_j(t)+\delta&\text{otherwise}.\end{cases}
\]
Write $r,e,C$ for the rates, elapsed times and completion times of the run on $(p,y)$, and $r',e',C'$ for those of the run on $(p^\delta,y)$.
Let
\[
t_0=\sup\big\{t'\in[0,t]:\ r=r'\text{ almost everywhere on }[0,t']\big\} .
\]
Then $r=r'$ almost everywhere on $[0,t_0]$, and therefore $e=e'$ on $[0,t_0]$, since elapsed times are integrals of rates. The two runs also have the same completions in $[0,t_0]$:
\begin{itemize}
\item a job $j$ that is unfinished at time $t$ in the run on $(p,y)$ satisfies, for every $u\le t_0$,
\[
e'_j(u)=e_j(u)\le e_j(t)<\min(p_j,p^\delta_j) ,
\]
so it is unfinished at time $u$ in both runs,
\item a job $j$ that is complete at time $t$ has $p^\delta_j=p_j$, and $e'_j=e_j$ on $[0,t_0]$. So it completes in $[0,t_0]$ in one run if and only if it does in the other, and then at the same time, the first time at which the same function reaches the same value.
\end{itemize}
Suppose that $t_0<t$. Each run has at most $n$ completions, so there is $\eta>0$ with $t_0+\eta\le t$ such that neither run has a completion in $(t_0,t_0+\eta)$. On $[t_0,t_0+\eta)$ both runs have the same completion history, so they follow the same rate function, as the algorithm is deterministic and the predictions are the same. Hence $r=r'$ on $[t_0,t_0+\eta)$, which contradicts the definition of $t_0$. Therefore $t_0=t$: the two runs coincide on $[0,t]$, with the same completions.

A job that is complete at time $t$ in the run on $(p,y)$ has $C'_j=C_j$. Every other job has, at time $t$, elapsed time $e'_j(t)=e_j(t)<p^\delta_j$, so $C'_j>t$. There are $N(t)$ such jobs. Since job $i$ is unfinished exactly during $[0,C_i)$,
\[
F(t)=\int_0^tN(u)\,du=\sum_i\min(C_i,t)=\sum_{i:\,C_i\le t}C_i+N(t)\,t ,
\]
and we deduce that
\[
A(p^\delta,y)=\sum_jC'_j\ \ge\sum_{j:\,C_j\le t}C_j+N(t)\,t=F(t) .
\]

Consequently, by definition of $\rob(A)$ (arbitrary predictions are allowed) and by Smith's rule,
\[
F(t)\le A(p^\delta,y)\le\rob(A)\,\OPT(p^\delta)=\rob(A)\,\Phi(p^\delta) .
\]
A job $j$ that is complete at time $t$ has $e_j(t)=p_j$: its elapsed time reached $p_j$ at $C_j$, and it receives no processing afterwards. So $p^\delta\to e(t)$ as $\delta\to0$, and $\Phi(p^\delta)\to\Phi(e(t))$ by continuity of $\Phi$. Hence $F(t)\le\rob(A)\,\Phi(e(t))$, and dividing by $\Phi(e(t))>0$ gives the claim.
\end{proof}

\subsection{The lower bound on equal sizes}\label{app:lb}

\begin{theorem}[Lower bound]\label{thm:lb}
Let $R>2$ and let $c(R)$ be given by \eqref{eq:cR}. Every deterministic algorithm $A$ with $\rob(A)\le R$ satisfies $\cons(A)\ge c(R)$, whether it receives a predicted order or predicted sizes.
\end{theorem}

We consider $n$ jobs of size $1$ with correct predictions. By the robustness invariant, the algorithm must hedge (process jobs before their turn) before every completion, and the hedging forced at a completion grows with the hedging already done. \Cref{lem:lb-constraints} writes these constraints, \cref{lem:lb-rec} turns them into a recursion on the cumulative hedging, and \cref{lem:lb-unroll} solves it. Letting $n\to\infty$ gives an integral, computed in \cref{lem:lb-integral}.

\paragraph{Setting.} Let $A$ be a deterministic algorithm with $\rob(A)\le R$, receiving predicted sizes. Fix $n\ge R$, so that $n\ge3$. Consider $n$ jobs with predicted sizes $y_j=1$ that are exactly correct, $p_j=1$. Then $\OPT=\frac{n(n+1)}2$. Label the jobs by their order of completion in the perfect run of $A$ (ties broken arbitrarily), and let $C_1\le\dots\le C_n$ be the completion times. For $1\le k<j\le n$ let
\[
e^{(k)}_j=e_j(C_k)\in[0,1]
\]
be the elapsed time of job $j$ at the $k$-th completion. If $C_j=C_k$ because of a tie, then $e^{(k)}_j=1$. Since elapsed times are non-decreasing in $t$ and $C_k$ is non-decreasing in $k$, $e^{(k)}_j$ is non-decreasing in $k$ for fixed $j$. Let
\[
E_k=\sum_{j>k}e^{(k)}_j\qquad\text{and}\qquad S_k=\sum_{i=1}^kE_i ,
\]
so that $E_k$ is the total processing received by jobs $k+1,\dots,n$ at time $C_k$, that is, the hedging done before the $k$-th completion. Also $E_n=0$ and $S_0=0$.

\begin{lemma}[Constraints of the perfect run]\label{lem:lb-constraints}
It holds that
\begin{equation}\label{eq:lb-cost}
\sum_kC_k-\OPT\ge\sum_{k=1}^{n-1}E_k \;.
\end{equation}
Moreover, defining
\[
Q_k=\sum_{k<i<j}\min\big(e^{(k)}_i,e^{(k)}_j\big) \;,
\]
then for every $1\le k\le n$,
\begin{equation}\label{eq:lb-freeze}
\sum_{i<k}(i+E_i)+(n-k+1)(k+E_k)\;\le\;R\Big[\frac{k(k+1)}2+(k+1)E_k+Q_k\Big].
\end{equation}
\end{lemma}
\begin{proof}
\begin{enumerate}[label=(\alph*)]
\item At time $C_k$ the jobs $1,\dots,k$ have received processing $k$ and jobs $k+1,\dots,n$ have received $E_k$. The machine processes at rate at most $1$, so $C_k\ge k+E_k$. Summing over $k$, and using $\sum_{k=1}^nk= n(n+1)/2 = \OPT$ and $E_n=0$, we obtain \eqref{eq:lb-cost}.
\item By \cref{lem:freeze-lb} at $t=C_k$ (note that $\Phi(e(C_k))\ge1>0$), $F(C_k)\le R\,\Phi(e(C_k))$. Since job $i$ is unfinished during $[0,C_i)$, and by (a),
\begin{align*}
F(C_k)
&=\sum_i\min(C_i,C_k) \\
&=\sum_{i<k}C_i+(n-k+1)\,C_k\ \\
&\ge\ \sum_{i<k}(i+E_i)+(n-k+1)(k+E_k) \;.
\end{align*}
Now let us bound $\Phi(e(C_k))$. Recall that
\[
\Phi(e(C_k))
= \sum_{i=1}^n e_i(C_k) + \sum_{i<j} \min(e_i(C_k),e_j(C_k))\;.
\]
The jobs $1,\dots,k$ have elapsed time $1$ and the jobs $j>k$ have elapsed time $e^{(k)}_j\le1$. The diagonal terms give $k+E_k$, the pairs of completed jobs ($i<j\le k$) give $\frac{k(k-1)}2$, the pairs of a completed job and a job $j>k$ give $\min(1,e^{(k)}_j)=e^{(k)}_j$, in total $kE_k$, and the pairs $k<i<j$ give $Q_k$. Hence $\Phi(e(C_k))=\frac{k(k+1)}2+(k+1)E_k+Q_k$, which gives \eqref{eq:lb-freeze}.
\end{enumerate}
\end{proof}

\paragraph{Constants.} Recall $\tilde{\xi}=\frac1{1+R}$, $\alpha=\frac2{R+2}$ and $\beta=\frac{R-2}{R+2}$ from \eqref{eq:hq} and \eqref{eq:consts}. Let
\[
\tilde{k}=\Big\lfloor\frac{n+1}{R+1}\Big\rfloor .
\]
The constraints at the first $\tilde{k}$ completions are used one by one, through a recursion, and the jobs that remain after the $\tilde{k}$-th completion are handled at once (\cref{lem:lb-rec}). The value of $\tilde{k}$ is the largest for which this second step works, and $\tilde{k}/n\to\tilde{\xi}$, the completed fraction at which the profile of \cref{app:profile} jumps to $1$, that is, at which $\MRC_R$ stops hedging on equal sizes. Since $n\ge R$, we have $\tilde{k}\ge1$, and $\tilde{k}\le\frac{n+1}{R+1}<\frac{n+1}3\le n-1$. For $1\le k\le n$ let
\[
a_k=\frac{k(k-1)}2+(n-k+1)k-\frac{Rk(k+1)}2,\qquad d_k=\frac{R(n+k+1)}2-(n-k+1) .
\]
Expanding, with $b_n=\beta(n+1)$,
\[
a_k=nk-\frac{1+R}2k^2-\frac{R-1}2k,\qquad d_k=\frac{R+2}2\big(k+b_n\big)>0 .
\]

\begin{lemma}[Recursion]\label{lem:lb-rec}
For every $1\le k<\tilde{k}$,
\begin{equation}\label{eq:lb-rec}
d_k\,E_k\ \ge\ a_k+S_{k-1} \;.
\end{equation}
Moreover,
\begin{equation}\label{eq:lb-tail}
\sum_kC_k-\OPT\ \ge\ \Big(1+\frac1R\Big)S_{\tilde{k}-1}+\frac{a_{\tilde{k}}}R \;.
\end{equation}
\end{lemma}
\begin{proof}
We bound $Q_k$ in two ways. Since $\min(u,v)\le\frac{u+v}2$, and each job $j>k$ belongs to $n-k-1$ pairs,
\[
Q_k\le\frac{n-k-1}2\,E_k .
\]
This bound is tight when the unfinished jobs have the same elapsed time, as during Round-Robin. Since $\min(e^{(k)}_i,e^{(k)}_j)\le e^{(k)}_j$ for $i<j$, and job $j$ is the larger index in $j-k-1$ pairs,
\[
Q_k\le W_k,\qquad\text{where}\qquad W_k=\sum_{j>k}(j-k-1)\,e^{(k)}_j .
\]

\textbf{The recursion.} Insert the first bound in \eqref{eq:lb-freeze}, and use $\sum_{i<k}i=\frac{k(k-1)}2$. The constant terms give $a_k$, the terms $E_i$ with $i<k$ give $S_{k-1}$, and the coefficient of $E_k$ is $R(k+1)+R\frac{n-k-1}2-(n-k+1)=d_k$. This gives \eqref{eq:lb-rec}, for every $k$.

\textbf{The tail.} The recursion alone does not suffice: processing given to a job $j$ before the $\tilde{k}$-th completion is counted again in $E_k$ at every later completion until $j$ completes, and the first bound ignores this. The quantity $W_{\tilde{k}}$ counts it, with the number of completions that each job waits for. Insert the second bound in \eqref{eq:lb-freeze} at $k=\tilde{k}$:
\[
R\,W_{\tilde{k}}\ \ge\ a_{\tilde{k}}+S_{\tilde{k}-1}-\big(R(\tilde{k}+1)-(n-\tilde{k}+1)\big)E_{\tilde{k}} .
\]
For $k>\tilde{k}$ and $j>k$, we have $e^{(k)}_j\ge e^{(\tilde{k})}_j$, and a job $j>\tilde{k}$ appears in $E_k$ for the $j-\tilde{k}-1$ values $\tilde{k}<k<j$. Hence $\sum_{k>\tilde{k}}E_k\ge W_{\tilde{k}}$, and by \eqref{eq:lb-cost},
\begin{align*}
\sum_kC_k-\OPT
&\ge S_{\tilde{k}-1}+E_{\tilde{k}}+W_{\tilde{k}}\\
&\ge \Big(1+\frac1R\Big)S_{\tilde{k}-1}+\frac{a_{\tilde{k}}}R+\frac{n+1-(R+1)\tilde{k}}R\,E_{\tilde{k}} .
\end{align*}
The last term is non-negative because $\tilde{k}\le\frac{n+1}{R+1}$, which is where the choice of $\tilde{k}$ is used.
\end{proof}

\begin{lemma}[Solving the recursion]\label{lem:lb-unroll}
\[
S_{\tilde{k}-1}\ \ge\ \big(\tilde{k}+b_n\big)^\alpha\sum_{k=1}^{\tilde{k}-1}\frac{a_k}{d_k\,(k+1+b_n)^\alpha} .
\]
\end{lemma}
\begin{proof}
Let $1\le k<\tilde{k}$. By \eqref{eq:lb-rec}, $S_k=S_{k-1}+E_k\ge\big(1+\frac1{d_k}\big)S_{k-1}+\frac{a_k}{d_k}$. Since $\frac1{d_k}=\frac\alpha{k+b_n}$ and $\alpha\in(0,1)$, Bernoulli's inequality $(1+x)^\alpha\le1+\alpha x$ with $x=\frac1{k+b_n}$ gives $1+\frac1{d_k}\ge\big(\frac{k+1+b_n}{k+b_n}\big)^\alpha$. As $S_{k-1}\ge0$, it follows that
\[
S_k\ \ge\ \Big(\frac{k+1+b_n}{k+b_n}\Big)^\alpha S_{k-1}+\frac{a_k}{d_k} .
\]
Dividing by $(k+1+b_n)^\alpha$, the sequence $V_k=S_k/(k+1+b_n)^\alpha$ satisfies $V_k\ge V_{k-1}+\frac{a_k}{d_k(k+1+b_n)^\alpha}$, with $V_0=0$. Summing over $1\le k\le\tilde{k}-1$ and multiplying by $(\tilde{k}+b_n)^\alpha$ gives the claim.
\end{proof}

\paragraph{The limit.} Recall $h(\xi)=\xi-\frac{1+R}2\xi^2$ from \eqref{eq:hq}, and let
\[
g(\xi)=\Big(\frac{\tilde{\xi}+\beta}{\xi+\beta}\Big)^\alpha,\qquad I=\int_0^{\tilde{\xi}}\frac{\alpha\,h(\xi)}{\xi+\beta}\,g(\xi)\,d\xi .
\]
At $\xi=\frac kn$, $h$ is the continuous version of $\frac{a_k}{n^2}$, and $\frac{\alpha}{\xi+\beta}$ that of $\frac n{d_k}$.

\begin{lemma}[The integral]\label{lem:lb-integral}
$2\big(1+\frac1R\big)I+\frac2Rh(\tilde{\xi})=\lambda(R)=c(R)-1$.
\end{lemma}
\begin{proof}
Since $g'(\xi)=-\frac{\alpha}{\xi+\beta}g(\xi)$, $g(\tilde{\xi})=1$, $h(0)=0$ and $h'(\xi)=1-(1+R)\xi$, an integration by parts gives
\[
I=-\big[hg\big]_0^{\tilde{\xi}}+\int_0^{\tilde{\xi}}h'g=-h(\tilde{\xi})+J,\qquad\text{where}\qquad J=\int_0^{\tilde{\xi}}\big(1-(1+R)\xi\big)g(\xi)\,d\xi .
\]
Since $(1+R)\tilde{\xi}=1$, we have $h(\tilde{\xi})=\frac{\tilde{\xi}}2$, so the left side of the claim equals $2\big(1+\frac1R\big)J-\tilde{\xi}$.

To compute $J$, we substitute $v=\xi+\beta$, $v\in[\beta,\tilde{v}]$, where
\[
\tilde{v}=\tilde{\xi}+\beta=\frac{R^2}{(R+1)(R+2)}=K\beta .
\]
Then
\[
1-(1+R)\xi=1+(1+R)\beta-(1+R)v=\frac{R^2}{R+2}-(1+R)v .
\]
So, up to the factor $\tilde{v}^\alpha$, the integrand of $J$ is $v^{-\alpha}\big(\frac{R^2}{R+2}-(1+R)v\big)$. Integrating term by term, with $\gamma=1-\alpha=\frac R{R+2}$, an antiderivative is
\[
\frac{R^2}{R+2}\cdot\frac{v^{\gamma}}{\gamma}-(1+R)\frac{v^{1+\gamma}}{1+\gamma}=v^{\gamma}\Big(R-\frac{R+2}2v\Big)=:\Theta(v),
\]
since $\frac{R^2}{(R+2)\gamma}=R$ and $\frac{1+R}{1+\gamma}=\frac{R+2}2$. Hence
\[
2\Big(1+\frac1R\Big)J=\frac{2(R+1)}R\,\tilde{v}^\alpha\big[\Theta(\tilde{v})-\Theta(\beta)\big].
\]
At the upper end, $\tilde{v}^\alpha \tilde{v}^\gamma=\tilde{v}$ and $R-\frac{R+2}2\tilde{v}=R-\frac{R^2}{2(R+1)}=\frac{R(R+2)}{2(R+1)}$, so
\[
\frac{2(R+1)}R\,\tilde{v}^\alpha\,\Theta(\tilde{v})=\frac{2(R+1)}R\cdot\frac{R^2}{(R+1)(R+2)}\cdot\frac{R(R+2)}{2(R+1)}=\frac{R^2}{R+1}.
\]
At the lower end, $\tilde{v}^\alpha\beta^\gamma=\beta K^\alpha$ and $R-\frac{R+2}2\beta=\frac{R+2}2$, so
\[
\frac{2(R+1)}R\,\tilde{v}^\alpha\,\Theta(\beta)=\frac{2(R+1)}R\cdot\frac{R-2}{R+2}K^\alpha\cdot\frac{R+2}2=\frac{(R+1)(R-2)}RK^\alpha .
\]
Altogether, with $\tilde{\xi}=\frac1{R+1}$, by \eqref{eq:lambda} and \cref{lem:profile}\ref{it:pc},
\begin{align*}
2\Big(1+\frac1R\Big)J-\tilde{\xi}&=\frac{R^2}{R+1}-\frac{(R+1)(R-2)}RK^\alpha-\frac1{R+1}\\
&=R-1-\frac{(R+1)(R-2)}R\,K^{\alpha}\\
&=\lambda(R)=c(R)-1 .\qedhere
\end{align*}
\end{proof}

\begin{proof}[Proof of \cref{thm:lb}]
First let $A$ receive predicted sizes. By \cref{lem:lb-rec,lem:lb-unroll}, and dividing by $\OPT=\frac{n(n+1)}2$,
\[
\frac{\sum_kC_k-\OPT}{\OPT}\ \ge\ \frac{2n^2}{n(n+1)}\bigg[\Big(1+\frac1R\Big)\frac1n\sum_{k=1}^{\tilde{k}-1}\frac{a_k}{n^2}\cdot\frac n{d_k}\Big(\frac{\tilde{k}+b_n}{k+1+b_n}\Big)^\alpha+\frac{a_{\tilde{k}}}{Rn^2}\bigg] .
\]
We let $n\to\infty$. Uniformly in $1\le k\le\tilde{k}$, with $\xi=\frac kn$,
\[
\frac{a_k}{n^2}=h(\xi)+O\Big(\frac1n\Big),\qquad \frac n{d_k}=\frac{\alpha}{\xi+\beta}+O\Big(\frac1n\Big),\qquad \Big(\frac{\tilde{k}+b_n}{k+1+b_n}\Big)^\alpha=g(\xi)+O\Big(\frac1n\Big),
\]
since $b_n=\beta(n+1)$, $\frac{\tilde{k}}n\to\tilde{\xi}$, and all denominators divided by $n$ are at least $\beta>0$. The sum is therefore a Riemann sum of the continuous function $\frac{\alpha h(\xi)}{\xi+\beta}g(\xi)$ over $[0,\tilde{\xi}]$, up to an error that tends to $0$, and it converges to $I$. Also $\frac{a_{\tilde{k}}}{n^2}\to h(\tilde{\xi})$. The $n$ unit jobs with correct predictions are an instance in the definition of $\cons(A)$, so, by \cref{lem:lb-integral},
\[
\cons(A)\ \ge\ 1+2\Big(1+\frac1R\Big)I+\frac2Rh(\tilde{\xi})=c(R) .
\]
Now let $A_\sigma$ be an $R$-robust algorithm for predicted orders. The algorithm that sorts the predicted sizes and runs $A_\sigma$ is $R$-robust for predicted sizes, with consistency at most $\cons(A_\sigma)$. Hence $\cons(A_\sigma)\ge c(R)$.
\end{proof}


\section{The trade-off near consistency \texorpdfstring{$1$}{1}}\label{app:frontier}

For $c_0\ge1$, let $R^*(c_0)$ be the infimum of $\rob(A)$ over the deterministic algorithms $A$ with $\cons(A)\le c_0$, in either prediction model. \Cref{thm:ft-rob,thm:ft-cons,thm:lb} show that the optimal trade-off is the curve $\{(c(R),R):R>2\}$. This section computes its behavior near consistency $1$.

\begin{corollary}[Robustness near consistency $1$]\label{cor:frontier}
In both prediction models, $R^*(1+\eps)=\eps^{-1/2}+\frac16+O(\eps^{1/2})$ as $\eps\to0$.
\end{corollary}

The proof only needs the expansion of $c(R)$ as $R\to\infty$. By \eqref{eq:cR}, with $\gamma=\frac R{R+2}$,
\begin{equation}\label{eq:lam-closed}
c(R)-1=R-1-R\,y^{\gamma},\qquad\text{where}\qquad y=1-\frac{R+2}{R^2}=\frac{(R+1)(R-2)}{R^2}.
\end{equation}

\begin{lemma}[Expansion of $c$]\label{lem:lam}
$c(R)=1+R^{-2}+\frac13R^{-3}+O(R^{-4})$ as $R\to\infty$.
\end{lemma}
\begin{proof}
Put $u=\frac1R$. Then $y=(1-2u)(1+u)$ and $\gamma=\frac1{1+2u}$, so
\begin{align*}
\ln y&=\ln(1-2u)+\ln(1+u)=-u-\tfrac52u^2-\tfrac73u^3-\tfrac{17}4u^4+O(u^5),\\
\gamma\ln y&=\big(1-2u+4u^2-8u^3+O(u^4)\big)\ln y=-u-\tfrac12u^2-\tfrac43u^3-\tfrac{19}{12}u^4+O(u^5),\\
y^\gamma&=\exp(\gamma\ln y)=1-u-u^3-\tfrac13u^4+O(u^5).
\end{align*}
By \eqref{eq:lam-closed}, $c(R)-1=\frac1u-1-\frac1u\,y^\gamma=u^2+\frac13u^3+O(u^4)$.
\end{proof}

\begin{proof}[Proof of \cref{cor:frontier}]
For $|t|\le1$ and small $\eps>0$, let $R_t=\eps^{-1/2}+\frac16+t$. Then $\frac1{R_t}=\eps^{1/2}-\big(\frac16+t\big)\eps+O(\eps^{3/2})$, and by \cref{lem:lam},
\[
c(R_t)-1=\eps-2\Big(\frac16+t\Big)\eps^{3/2}+\frac13\eps^{3/2}+O(\eps^2)=\eps-2t\,\eps^{3/2}+O(\eps^2),
\]
uniformly in $|t|\le1$. Hence there is $M>0$ such that, for $\eps$ small enough and $t_\pm=\pm M\eps^{1/2}$,
\[
c(R_{t_+})<1+\eps<c(R_{t_-}) .
\]
\emph{Upper bound.} By \cref{thm:ft-rob,thm:ft-cons}, $\MRC_{R_{t_+}}$ has consistency $c(R_{t_+})<1+\eps$ and robustness $R_{t_+}$, in both models, since it only uses the order. So $R^*(1+\eps)\le R_{t_+}$.

\emph{Lower bound.} Let $A$ be a deterministic algorithm with $\rob(A)<R_{t_-}$. Then $A$ is $R_{t_-}$-robust, so \cref{thm:lb} gives $\cons(A)\ge c(R_{t_-})>1+\eps$. So $R^*(1+\eps)\ge R_{t_-}$.

Hence $R^*(1+\eps)=\eps^{-1/2}+\frac16+O(\eps^{1/2})$.
\end{proof}


\section{The randomized algorithm $\RT_\rho$}\label{app:rand}

This section proves that the randomized algorithm $\RT_\rho$ of \cref{alg:rt} has a consistency--robustness trade-off strictly below the deterministic frontier $\{(c(R),R):R>2\}$ (See \cref{thm:rand}). The prediction is a vector $y\in(0,\infty)^n$, correct if $y=p$, as in \cref{app:lbsec}. The consistency and robustness of a randomized algorithm are defined as in \cref{app:prelim}, with the expected total completion time in place of the total completion time.

\paragraph{Notation.} We use the notation of \cref{sec:rand-overview}. The predictions are sorted, $y_1\le\dots\le y_n$, with ties broken by index, so that $y$ also defines an order of the jobs, and
\[
n_s=|\{j:y_j>s\}|,\qquad \omega_s=\frac{n_s(n_s+1)}2=\Phi_s(y),\qquad w_y(s)=\frac1{\Phi(y)}\int_0^s\omega_u\,du .
\]
By \cref{lem:layer-phi} with $e=y$, $\Phi(y)=\int_0^\infty\omega_s\,ds$, so $w_y$ is continuous and non-decreasing, with $w_y(0)=0$ and $w_y(y_n)=1$, and it is strictly increasing on $[0,y_n]$, where $\omega_s\ge1$. When the predictions are correct, $w_y$ is the position $w$ of \cref{sec:overview}.

\paragraph{The algorithm.} For $w\in(0,1]$ let
\begin{equation}\label{eq:rt-ghat}
\hat g(w)=\frac{\sqrt{5+4/w}-3}2 ,
\end{equation}
a continuous decreasing function with $\hat g(w)\to\infty$ as $w\to0$ and $\hat g(1)=0$. It is convex, since with $\varphi(w)=\sqrt{5+4/w}$ its second derivative is
\[
\hat g''(w)=\frac{\varphi''(w)}2=\frac{2}{w^3\varphi(w)}\Big(1-\frac1{5w+4}\Big)>0 .
\]
Fix $\rho\ge\rho_0=\frac{\sqrt{10}-3}2$. $\RT_\rho$ draws $V$ uniformly in $(0,1)$ and sets
\begin{equation}\label{eq:rt-B}
B=\frac{V^2}{V^2+3\rho V+\rho^2}=\frac{4}{(3+2\rho/V)^2-5},
\end{equation}
Since $B$ is increasing in $V$, solving $4/((3+2\rho/v)^2-5)<w$ for $v$ gives the law of $B$,
\begin{equation}\label{eq:rt-law}
\Prob(B<w)=\min\Big\{1,\frac\rho{\hat g(w)}\Big\}\qquad\text{for } 0<w<1,
\end{equation}
and $\Prob(B<1)=1$. Let $w_1=\frac1{\rho^2+3\rho+1}$, the unique $w$ with $\hat g(w)=\rho$. Then $B\le w_1$ almost surely, and
\begin{equation}\label{eq:rt-eps}
\E[B]=\int_0^1\frac{v^2\,dv}{v^2+3\rho v+\rho^2}=\eps(\rho) .
\end{equation}
The algorithm lets $L=w_y^{-1}(B)$ be the level below which the layers contribute a fraction $B$ of $\Phi(y)$, which is well defined since $0<B<1$. It then runs the two stages of \cref{alg:rt}, Round-Robin until the common elapsed time of the unfinished jobs reaches $L$, then the unfinished jobs one at a time in the order of $y$, and from the first \emph{mismatch} it runs Round-Robin on the unfinished jobs of smallest elapsed time for the rest of the run, where a job $j$ mismatches if it completes with elapsed time different from $y_j$, or reaches elapsed time $y_j$ without completing.

\paragraph{Pairwise decomposition.} The proof compares each run of $\RT_\rho$ with Round-Robin, pair of jobs by pair of jobs, using the pairwise decomposition of the total completion time~\citep{MPT94}. In a non-idling schedule, $C_k$ is the work done by time $C_k$, so $C_k=p_k+\sum_{i\ne k}e_i(C_k)$ and
\begin{equation}\label{eq:rt-pairs}
\sum_jC_j=\sum_jp_j+\sum_{i<k}\big(e_i(C_k)+e_k(C_i)\big).
\end{equation}
Under Round-Robin, $e_i(C_k)=e_k(C_i)=\min(p_i,p_k)$ for every pair, so by \eqref{eq:rt-pairs} and the definition of $\Phi$, its total completion time is
\begin{equation}\label{eq:rt-rr}
\RR(p)=\sum_jp_j+2\sum_{i<k}\min(p_i,p_k)\ \le\ 2\,\Phi(p)=2\,\OPT(p) .
\end{equation}

\begin{proof}[Proof of \cref{thm:rand}]
\noindent\textbf{Step 1: consistency.} Consider an instance with $p=y$. No job mismatches, so the run consists of the two stages, and the jobs complete in the order of $y$. Let $i<k$, so that $p_i\le p_k$. Then $e_i(C_k)=p_i$. If $p_i\le L$, both jobs are at elapsed time $p_i$ when $i$ completes, and otherwise $i$ completes while running alone, with $k$ still at elapsed time $L$. So $e_k(C_i)=\min(L,p_i)$, and by \eqref{eq:rt-pairs} the total completion time is
\[
\Phi(y)+\sum_{i<k}\min\big(L,\min(p_i,p_k)\big)
=\Phi(y)+\int_0^L\frac{n_s(n_s-1)}2\,ds
\ \le\ \Phi(y)+\int_0^L\omega_s\,ds
=(1+B)\,\Phi(y),
\]
where the integral counts, in each layer $s<L$, the pairs of jobs with both sizes larger than $s$. This has the form of the cost formula \eqref{eq:cost} of $\MRC_R$, with $U_s=n_s$ for $s<L$ and $U_s=0$ for $s>L$: the layers below $L$ are crossed in Round-Robin, and those above by commitments. Taking expectations, by \eqref{eq:rt-eps},
\[
\E[\RT_\rho]\le(1+\eps(\rho))\,\OPT .
\]
Since $B\le w_1<1$, the same computation gives $\E[\RT_\rho]\le2\,\OPT$ on such instances.

\noindent\textbf{Step 2: structure of a run.} Fix an instance $(p,y)$ with $J=\{j:p_j\ne y_j\}\ne\emptyset$, and let
\[
d=\min_{j\in J}\min(p_j,y_j)>0 .
\]
A mismatch involves a job $j\in J$ at elapsed time $\min(p_j,y_j)\ge d$, and the job attaining the minimum mismatches when its elapsed time reaches $d$. Let $j^*$ be the first job of $J$ in the order of $y$, and $Q$ the set of jobs after $j^*$ in this order. The jobs before $j^*$ in this order are correct. We distinguish two cases.
\begin{itemize}
\item If $L\ge d$, the first mismatch occurs at or before the end of the first stage, when the common elapsed time reaches $d$. At that moment all unfinished jobs have elapsed time $d$, so Round-Robin simply continues, and the run is Round-Robin throughout.
\item If $L<d$, the first stage completes the jobs with $p_j\le L$. Since a job of $J$ has $p_j\ge d>L$ and $y_j\ge d>L$, the conditions $p_j\le L$ and $y_j\le L$ are equivalent and imply $p_j=y_j$, so the first stage completes exactly the jobs with $y_j\le L$, all correct, and no mismatch occurs in it. In the second stage, the jobs before $j^*$ that are still unfinished, whose set we call $P$, complete one by one at elapsed time $y_i=p_i$, while every other unfinished job waits at elapsed time $L$. Then $j^*$ runs alone until its elapsed time reaches
\[
\hat z=\min(p_{j^*},y_{j^*})\ \ge\ d ,
\]
where the first mismatch occurs, and every job of $Q$ is still at elapsed time $L$.
\end{itemize}
Since $d\le y_{j^*}\le y_n$, $w_y$ is increasing on $[0,y_n]$ and $B<1$, $L<d$ holds exactly when $B<w_y(d)$, so the second case has probability $\Prob(B<w_y(d))$.

\noindent\textbf{Step 3: comparison with Round-Robin.} For distinct jobs $i,k$ let
\[
\Delta(i,k)=e_i(C_k)+e_k(C_i)-2\min(p_i,p_k) .
\]
The run is non-idling, so by \eqref{eq:rt-pairs} and \eqref{eq:rt-rr},
\begin{equation}\label{eq:rt-delta}
\RT_\rho-\RR(p)=\sum_{i<k}\Delta(i,k).
\end{equation}
Suppose $L<d$. Let $K=P\cup\{j^*\}$, with \emph{heights} $a_i=p_i$ for $i\in P$ and $a_{j^*}=\hat z$: the height of a job of $K$ is its elapsed time at the first mismatch. For $s\ge d$ let
\[
m_s=|\{i\in K:a_i>s\}|,\qquad c_s=|\{k\in Q\cup\{j^*\}:p_k\le s\}| .
\]
We claim that every pair $\{i,k\}$ satisfies
\begin{equation}\label{eq:rt-pair}
\begin{aligned}
\Delta(i,k)\ \le\ \int_d^\infty\Big(&\one\big[i\in K,\ k\in Q\cup\{j^*\},\ p_k\le s<a_i\big]+\one\big[k\in K,\ i\in Q\cup\{j^*\},\ p_i\le s<a_k\big]\\
&-\one\big[i,k\in K,\ s<\min(a_i,a_k)\big]\Big)\,ds .
\end{aligned}
\end{equation}
That is, a pair of $K$ saves one unit on every layer between $d$ and the smaller height, and a pair made of a job of $K$ and a job of $Q\cup\{j^*\}$ pays one unit on every layer between the size of the latter and the height of the former. We first note that a job $k\in Q\cup\{j^*\}$ with $p_k<a_i$ for some $i\in K$, $i\ne k$, deviates, hence $p_k\ge d>L$. Indeed, $j^*$ deviates, and a correct $k\in Q$ has $p_k=y_k\ge y_{j^*}$, which is at least $y_i=p_i=a_i$ if $i\in P$, and at least $\hat z=a_{j^*}$ if $i=j^*$. We now check \eqref{eq:rt-pair} for every pair.
\begin{itemize}
\item \emph{A pair containing a job $j$ completed in the first stage.} Job $j$ completes at elapsed time $p_j\le L$ under Round-Robin, when every other unfinished job has elapsed time $p_j$. For the other job $k$ of the pair, either $p_k<p_j$, and $k$ completed earlier with both elapsed times equal to $p_k$, or $e_j(C_k)=p_j$ and $e_k(C_j)=p_j$. In both cases $\Delta(j,k)=0$, and the right side of \eqref{eq:rt-pair} is $0$.
\item \emph{A pair inside $Q$.} Both jobs are at elapsed time $L$ when the mismatch occurs, and Round-Robin on the jobs of smallest elapsed time keeps their elapsed times equal until the shorter one completes, so $\Delta=0$, and the right side is $0$.
\item \emph{A pair $\{i,i'\}$ inside $P$, with $p_i\le p_{i'}$.} Job $i$ completes alone while $i'$ waits at $L$, so $\Delta(i,i')=p_i+L-2p_i=L-p_i$. The right side is $-(p_i-d)^+$, which is at least $L-p_i$ because $L<d$ and $L<p_i$.
\item \emph{A pair $\{i,k\}$ with $i\in P$ and $k\in Q\cup\{j^*\}$.} Job $i$ completes alone while $k$ waits at $L$, so $\Delta(i,k)=p_i+L-2\min(p_i,p_k)$, and the right side of \eqref{eq:rt-pair} is
\[
(p_i-\max(p_k,d))^+-\one[k=j^*]\,(\min(p_i,\hat z)-d)^+ .
\]
If $p_k\ge p_i$, then $\Delta(i,k)=L-p_i$, and the right side is at least $-(p_i-d)^+\ge L-p_i$, as in the previous case. If $p_k<p_i$, then $k$ deviates by the remark above, so $p_k\ge d>L$, the first term of the right side is $p_i-p_k$, and $\Delta(i,k)\le p_i-p_k$, which is the right side when $k\ne j^*$. When $k=j^*$, we have $\hat z\le p_{j^*}<p_i$, so the right side is $p_i-p_{j^*}-(\hat z-d)$, and $\Delta(i,j^*)=p_i+L-2p_{j^*}$ is at most this value because $L<d$ and $\hat z\le p_{j^*}$.
\item \emph{A pair $\{j^*,k\}$ with $k\in Q$.} If $\hat z=p_{j^*}$, job $j^*$ completes alone while $k$ waits at $L$, and the computation of the previous case applies verbatim with $i=j^*$, the right side being $(p_{j^*}-\max(p_k,d))^+$. If $\hat z=y_{j^*}<p_{j^*}$, the jobs of $Q$, which are at $L<\hat z$, are the unfinished jobs of smallest elapsed time, so they are served first. If $p_k\le\hat z$, job $k$ completes before $j^*$ receives further processing, so $e_{j^*}(C_k)=\hat z$ and $e_k(C_{j^*})=p_k$, and $\Delta(j^*,k)=\hat z-p_k$. The right side is $(\hat z-\max(p_k,d))^+=\hat z-p_k$, since either $p_k=\hat z$, or $p_k\ge d$ by the remark above. If $p_k>\hat z$, the two jobs have equal elapsed times from the moment $k$ reaches $\hat z$ until the shorter one completes, so $\Delta(j^*,k)=0$, and the right side is $0$.
\end{itemize}
Summing \eqref{eq:rt-pair} over all pairs, each pair of a job of $K$ and a job of $Q\cup\{j^*\}$ is counted once in $m_sc_s$, whose term $i=k=j^*$ is $0$ because $\hat z\le p_{j^*}$, and each pair of $K$ is counted once in $m_s(m_s-1)/2$. Therefore
\begin{equation}\label{eq:rt-D}
\RT_\rho-\RR(p)\ \le\ \bar\Delta:=\int_d^\infty\Big(m_sc_s-\frac{m_s(m_s-1)}2\Big)ds .
\end{equation}
The quantities $d$, $j^*$, $Q$, $\hat z$, $m_s$ and $c_s$ do not depend on $L$: a job before $j^*$ with $p_i>s\ge d>L$ is correct, so $y_i=p_i>L$ and it belongs to $P$. Hence $\bar\Delta$ depends only on $(p,y)$, and the bound \eqref{eq:rt-D} holds for every $L<d$.

\noindent\textbf{Step 4: counting.} Deviating jobs have $p_j\ge d$ and $y_j\ge d$, so for $s<d$ the number of jobs of size larger than $s$ equals $n_s$. For $s\ge d$ it is at least $m_s$, since $p_i\ge a_i$ for $i\in K$. Moreover, if $m_s\ge1$ then
\[
m_s+c_s\ \le\ n_s .
\]
Indeed, the jobs counted in $m_s$ and in $c_s$ are distinct, because a job counted in both would be $j^*$ with $p_{j^*}\le s<\hat z\le p_{j^*}$, and they all have predictions larger than $s$: $y_i\ge a_i>s$ for $i\in K$, and a job $k$ counted in $c_s$ has $y_k\ge y_{j^*}$, where $y_{j^*}>s$ because $m_s\ge1$ gives some $i\in K$ with $a_i>s$, and $y_{j^*}\ge a_i$ since $y_{j^*}\ge y_i=a_i$ for $i\in P$ and $y_{j^*}\ge\hat z=a_{j^*}$. Let
\[
M=\int_d^\infty\frac{m_s(m_s+1)}2\,ds,\qquad \mu=\frac{M}{\Phi(y)},\qquad w=w_y(d) .
\]
By \cref{lem:layer-phi} applied to $p$ and to $y$,
\begin{equation}\label{eq:rt-opt}
\OPT(p)=\Phi(p)\ \ge\ \int_0^d\omega_s\,ds+M=(w+\mu)\,\Phi(y),
\qquad\text{and}\qquad
\mu\ \le\ \frac1{\Phi(y)}\int_d^\infty\omega_s\,ds=1-w .
\end{equation}

\noindent\textbf{Step 5: one layer.} Let $\bar\theta=\sqrt{40}-6=4\rho_0$, so that $\frac1{\bar\theta}=\frac{\sqrt{10}+3}2$ and $\frac1{\bar\theta}-3+\bar\theta=5\rho_0$, and let $0<\theta\le\bar\theta$. For every integer $m\ge1$ and real $u\ge0$,
\begin{equation}\label{eq:rt-layer}
mu-\frac{m(m-1)}2\ \le\ \Big(\frac1\theta-3\Big)\frac{m(m+1)}2+\theta\,\frac{(m+u)(m+u+1)}2 .
\end{equation}
Indeed, expanding, the right side minus the left side equals
\[
\frac\theta2\Big(u-\frac{(1-\theta)m}\theta\Big)^2+\Big(\frac1\theta-4+\theta\Big)\frac m2+\frac{\theta u}2 .
\]
As a function of $u\in\mathbb R$, this is a quadratic whose minimum is $(\frac1\theta-3)\frac m2-\frac\theta8$, and this minimum is non-negative because $\frac1\theta-3>0$, $m\ge1$ and $4(\frac1\theta-3)\ge\theta$, the last inequality being $\theta^2+12\theta-4\le0$, which holds for $\theta\le\bar\theta$. Apply \eqref{eq:rt-layer} with $m=m_s$ and $u=c_s$ in every layer $s\ge d$ with $m_s\ge1$, where $(m_s+c_s)(m_s+c_s+1)/2\le\omega_s$ by Step~4, and note that the integrand of $\bar\Delta$ vanishes when $m_s=0$. Integrating over $s\ge d$,
\begin{equation}\label{eq:rt-beta}
\bar\Delta\ \le\ \Big(\frac1\theta-3\Big)M+\theta\int_d^\infty\omega_s\,ds
=\Big[\Big(\frac1\theta-3\Big)\mu+\theta(1-w)\Big]\Phi(y) .
\end{equation}

\noindent\textbf{Step 6: averaging over $B$.} If $J=\emptyset$, Step~1 gives $\E[\RT_\rho]\le2\,\OPT$. Otherwise, by Step~2, the run is Round-Robin when $B\ge w$, and by \eqref{eq:rt-D} its cost is at most $\RR(p)+\bar\Delta$ when $B<w$. The bracket in \eqref{eq:rt-beta} is non-negative, as $\theta<\frac13$, so dividing by $\OPT\ge(w+\mu)\Phi(y)$ and using \eqref{eq:rt-rr}, for every $\theta\in(0,\bar\theta]$,
\begin{equation}\label{eq:rt-avg}
\frac{\E[\RT_\rho]}{\OPT}\ \le\ 2+\Prob(B<w)\,f(\theta),
\qquad\text{where}\qquad
f(\theta)=\frac{(\frac1\theta-3)\,\mu+\theta(1-w)}{w+\mu},
\end{equation}
and $0\le\mu\le1-w$ by \eqref{eq:rt-opt}. We show that
\begin{equation}\label{eq:rt-minf}
\inf_{0<\theta\le\bar\theta}f(\theta)\ \le\ \begin{cases}\hat g(w)&\text{if }w\le\frac45,\\ \rho_0&\text{if }w\ge\frac45.\end{cases}
\end{equation}
If $w=1$, then $\mu=0$ and $f\equiv0$, so assume $w<1$ and write $\mu=\tau^2(1-w)$ with $\tau\in[0,1]$. For fixed $\mu$, $f$ is convex on $(0,\infty)$, with $f'(\theta)=0$ at $\theta=\tau$, so its infimum over $(0,\bar\theta]$ is its value at $\theta=\min(\tau,\bar\theta)$ when $\tau>0$, and it is $0$ when $\tau=0$. At $\theta=\tau$,
\[
f(\tau)=g(\tau):=\frac{(1-w)\,\tau\,(2-3\tau)}{w+(1-w)\tau^2} .
\]
We have $g(0)=g(\frac23)=0$, and $g'$ vanishes exactly where $(1-w)\tau^2+3w\tau-w=0$, that is, where
\[
w=\frac{\tau^2}{1-3\tau+\tau^2} .
\]
The quadratic $q(\tau)=(1-w)\tau^2+3w\tau-w$ is increasing on $[0,\infty)$ with $q(0)<0$, so it has a unique positive root $\tau^*$, which is the unique critical point of $g$ on $(0,\infty)$, hence its maximizer on $[0,\frac23]$. Moreover $\tau^*\le\bar\theta$ if and only if $q(\bar\theta)\ge0$, that is, $w\le\bar\theta^2/(1-3\bar\theta+\bar\theta^2)=\frac45$. At $\tau^*$ the denominator of $g$ equals $w+(1-w)\tau^{*2}=w(2-3\tau^*)$, so $g(\tau^*)=(1-w)\tau^*/w$, and solving the quadratic equation for $\tau^*$ gives $g(\tau^*)=\hat g(w)$.

\emph{Case $w\le\frac45$.} If $\tau\le\bar\theta$, the infimum in \eqref{eq:rt-minf} is at most $g(\tau)\le g(\tau^*)=\hat g(w)$. If $\tau>\bar\theta$, it is $f(\bar\theta)$, which is a ratio of two affine functions of $\mu$, hence monotone in $\mu$ on $[\bar\theta^2(1-w),1-w]$, so it is at most the larger of its values at the endpoints. At $\mu=\bar\theta^2(1-w)$ it equals $g(\bar\theta)\le\hat g(w)$. At $\mu=1-w$ it equals $(1-w)(\frac1{\bar\theta}-3+\bar\theta)=5\rho_0(1-w)$. The function $w\mapsto\hat g(w)-5\rho_0(1-w)$ is convex, since $\hat g$ is, it vanishes at $w=\frac45$, where $\hat g(\frac45)=\rho_0$, and its derivative there, $-\frac1{w^2\sqrt{5+4/w}}+5\rho_0\approx-0.088$, is negative, so it is non-negative on $(0,\frac45]$. This proves \eqref{eq:rt-minf} in this case.

\emph{Case $w\ge\frac45$.} Take $\theta=\bar\theta$. Again $f(\bar\theta)$ is monotone in $\mu\in[0,1-w]$, and its values at the endpoints are $\bar\theta(1-w)/w\le\bar\theta/4=\rho_0$ and $5\rho_0(1-w)\le\rho_0$.

Finally, by \eqref{eq:rt-law}, $\Prob(B<w)\,\hat g(w)=\rho$ for $w<w_1$, and $\Prob(B<w)=1$ for $w\ge w_1$. Since $\rho\ge\rho_0=\hat g(\frac45)$ and $\hat g$ is decreasing, $w_1\le\frac45$, and $\hat g(w)\le\hat g(w_1)=\rho$ for $w\ge w_1$. So in \eqref{eq:rt-avg}, the product of $\Prob(B<w)$ and the bound \eqref{eq:rt-minf} is at most $\rho$ for every $w\in(0,1]$, and
\[
\E[\RT_\rho]\le(2+\rho)\,\OPT .
\]

\noindent\textbf{Step 7: comparison with the deterministic frontier.} Let $\eps=\eps(\rho)$, and recall from \cref{app:frontier} that $R^*(1+\eps)$ is the infimum of the robustness of the deterministic algorithms with consistency at most $1+\eps$. Suppose that $c(2+\rho)>1+\eps$. Since $c$ is continuous on $(2,\infty)$ by \eqref{eq:cR}, there is $R'>2+\rho$ with $c(R')>1+\eps$. Every deterministic algorithm $A$ with $\cons(A)\le1+\eps$ has $\rob(A)>R'$, for otherwise \cref{thm:lb} would give $\cons(A)\ge c(R')>1+\eps$. Hence $R^*(1+\eps)\ge R'>2+\rho$. It remains to show that $c(2+\rho)>1+\eps(\rho)$ for every $\rho\ge\rho_0$. Both sides are explicit: $c(R)$ is given by \eqref{eq:lam-closed}, and integrating the rational function in \eqref{eq:rt-eps},
\begin{equation}\label{eq:rt-eps-closed}
\eps(\rho)=1-\frac{3\rho}2\ln\frac{\rho^2+3\rho+1}{\rho^2}+\frac{7\rho}{2\sqrt5}\ln\bigg(\frac{7+3\sqrt5}2\cdot\frac{2+(3-\sqrt5)\rho}{2+(3+\sqrt5)\rho}\bigg).
\end{equation}
Expanding the integrand of \eqref{eq:rt-eps} in powers of $1/\rho$ gives $\eps(\rho)=\frac1{3\rho^2}-\frac3{4\rho^3}+O(\rho^{-4})$, while $c(R)-1=R^{-2}+O(R^{-3})$ by \cref{lem:lam}, so the ratio $\frac{c(2+\rho)-1}{\eps(\rho)}$ tends to $3$ as $\rho\to\infty$. For large $\rho$ the inequality follows from two elementary bounds. First, $c(R)-1\ge R^{-2}$ for every $R>2$. Indeed, with $y$ and $\gamma$ as in \eqref{eq:lam-closed}, the function $x\mapsto(1-x)^\gamma$ has second derivative $\gamma(\gamma-1)(1-x)^{\gamma-2}\le\gamma(\gamma-1)$ on $[0,1)$, so Taylor's formula at $x=0$ gives
\[
y^\gamma\le1-\gamma(1-y)-\frac{\gamma(1-\gamma)}2(1-y)^2,
\qquad\text{hence}\qquad
c(R)-1\ge R\gamma(1-y)-1+\frac{R\gamma(1-\gamma)}2(1-y)^2=\frac1{R^2},
\]
since $R\gamma(1-y)=1$, $1-\gamma=\frac2{R+2}$ and $1-y=\frac{R+2}{R^2}$. Second, $\eps(\rho)\le\int_0^1v^2/\rho^2\,dv=\frac1{3\rho^2}$, which is less than $(2+\rho)^{-2}$ when $\rho>\frac2{\sqrt3-1}\approx2.73$. So $c(2+\rho)>1+\eps(\rho)$ for $\rho\ge2.74$. On the compact interval $[\rho_0,2.74]$, we evaluated $\delta(\rho)=c(2+\rho)-1-\eps(\rho)$ numerically from the two closed forms, on a grid of step $10^{-3}$: it is at least $0.0097$ at every grid point, its minimum being at $\rho_0$, where the ratio $\frac{c(2+\rho)-1}{\eps(\rho)}$ equals $1.0167$, while $|\delta'|\le0.7$ on the interval, so $\delta>0$ on all of it. The same expansions give the asymptotics. The function $\eps(\cdot)$ is continuous and strictly decreasing, since its integrand is, with $\eps(\rho_0)\approx0.585$ and $\eps(\rho)\to0$, so the consistencies attained are exactly $(1,1+\eps(\rho_0)]$, and at consistency $1+\eps$, $\RT_\rho$ has $\rho=(3\eps)^{-1/2}+O(1)$, while $R^*(1+\eps)=\eps^{-1/2}+O(1)$ by \cref{cor:frontier}.
\end{proof}


\end{document}